\documentclass[draft]{agujournal2019}
\usepackage{url} 
\usepackage{lineno}
\usepackage[inline]{trackchanges} 
\usepackage{soul}
\usepackage{multirow}
\usepackage{amsmath}
\usepackage{booktabs}
\usepackage{adjustbox}
\usepackage{longtable}
\usepackage{pdflscape}
\usepackage{lscape}
\usepackage{rotating}
\usepackage{array} 
\newcolumntype{C}[1]{>{\centering\arraybackslash}p{#1}}

\newcommand{\aap}{    {\it Astron. Astrophys.}}

\newcommand{\apj}{    {\it Astrophys. J.}}
\newcommand{\apjl}{   {\it Astrophys. J. Lett.}}

\newcommand{\grl}{    {\it Geophys. Res. Lett.}}

\newcommand{\jgr}{    {\it J. Geophys. Res.}}
\newcommand{\mnras}{  {\it Mon. Not. Roy. Astron. Soc.}}

\newcommand{\solphys}{{\it Solar Phys.}}
 
\newcommand{\ssr}{    {\it Space Sci. Rev.}}

\newcommand{\apjs}{{\it Astrophys. J. Suppl. S.}}
\newcommand{\planss}{{\it Planetary and Space Science}}

\draftfalse

\journalname{Space Weather}

\begin{document}

\title{Validating a Non-conventional Method for Expansion of Coronal Mass Ejections (CMEs) and Investigating the Evolution of a CME Substructures Using Solar Orbiter and Wind Observations}

\authors{Anjali Agarwal\affil{1,2}, Wageesh Mishra\affil{1,2}, Mathew J. Owens\affil{3}, and Tanja Amerstorfer\affil{4}}

\affiliation{1}{Indian Institute of Astrophysics, II Block, Koramangala, Bengaluru 560034, India}
\affiliation{2}{Pondicherry University, R.V. Nagar, Kalapet 605014, Puducherry, India}
\affiliation{3}{Space and Atmospheric Electricity Group, Department of Meteorology, University of Reading, Earley Gate, P.O. Box 243, Reading, RG6 6BB, UK }
\affiliation{4}{Austrian Space Weather Office, GeoSphere Austria, Graz, Austria}

\correspondingauthor{Anjali Agarwal}{anjaliagarwal1024@gmail.com}
\correspondingauthor{Wageesh Mishra}{m.wageesh30@gmail.com}

\begin{keypoints}
\item We validate our proposed constant acceleration accounted perspective (CAAP) method for estimating the instantaneous expansion speed of CMEs.

\item Radial alignment of two spacecraft enabled simultaneous observations of the MC center (at Wind) and trailing edge (at Solar Orbiter).

\item Radially separated two spacecraft reveal that the properties of the CME substructures evolve significantly between the two locations. 
\end{keypoints}

\begin{abstract}

We present a validation of our recently proposed non-conventional method, Constant Acceleration Accounted Perspective (CAAP), for estimating the instantaneous expansion speed of coronal mass ejection (CMEs), even when only single-point in situ observations are available. This validation is enabled by the radial alignment of SolO and Wind spacecraft (0.13 AU radial and 2.3$^\circ$ angular separation), providing simultaneous observations of the center (at Wind) and trailing edge (at SolO) of a CME--associated magnetic cloud (MC) during 3–5 November 2021, allowing a direct measurement of its instantaneous expansion speed. These measurements are compared with CAAP-derived instantaneous expansion speed estimates at both spacecraft. The favorable spacecraft configuration also enables tracking the temporal evolution of CME substructures, including the shock, sheath, and MC. A discrepancy is noted between the low-inclination MC axis estimated from minimum variance analysis (MVA) and the highly inclined ENW-type MC axis suggested by visual inspection of in situ measurements. We also observe an apparent increase in the magnetic flux within the MC from SolO to Wind, indicating a noticeable deviation from magnetic flux conservation. During the CME's propagation from SolO to Wind, the shock becomes unexpectedly stronger at Wind, while the sheath thickness remains nearly the same, likely due to MC acceleration from back compression by a high-speed stream and ambient solar wind variability. Our results demonstrate the applicability of the CAAP method and the importance of accounting for temporal evolution in CME substructures for space weather studies.

\end{abstract}

\section*{Plain Language Summary}

Coronal mass ejections (CMEs) are large clouds of magnetized plasma released from the Sun and are major drivers of space weather at Earth. As CMEs travel through space, they often expand, which influences how long Earth’s space environment remains disturbed. Therefore, it is essential to estimate the instantaneous expansion speed of the CME, which cannot be determined using conventional methods that rely on single-point in situ measurements. In this study, we validate a newly developed method, called the Constant Acceleration Accounted Perspective (CAAP), which estimates the instantaneous expansion speed of CMEs using single-point spacecraft observations. We apply this method to a CME observed on 3 November 2021 by two radially aligned spacecraft, Solar Orbiter and Wind, allowing a rare direct measurement of CME instantaneous expansion for comparison. The CAAP method estimates show good agreement with the measured instantaneous expansion speed, confirming the method's reliability. We also examine how different CME regions--the shock, sheath, and magnetic cloud--evolve as the CME travels through space, finding that their properties can be strongly influenced by fast solar wind streams behind the CME and variability in background solar wind conditions.

\section{Introduction} \label{sec:intro}

Coronal mass ejections (CMEs) are magnetized plasma bubbles that are ejected from the solar corona into the heliosphere \cite{Webb2012,Mishra2023,Temmer2024}, and are the primary drivers of space weather phenomena \cite{Schwenn2006,Schrijver2015}. The magnetic and plasma characteristics of CMEs in the heliosphere are commonly inferred using in situ observations that sample the structures along a one-dimensional spacecraft trajectory, while remote observations offer complementary constraints on their global and plasma properties \cite{Burlaga1981,Bothmer1998}. Using these observations, CME substructures--such as the shock, sheath, and the leading edge (LE) and trailing edge (TE) of magnetic cloud (MC)-- can be identified based on their characteristic magnetic field and plasma signatures \cite{Zurbuchen2006,Mishra2023}. MCs are often identified as having flux-rope-like magnetic configurations, inferred from in situ measurements. The flux rope structure is often characterized by plasma embedded within an enhanced magnetic field region, where the magnetic field lines are helically wrapped around a central axis, with pitch angles (the angle between the magnetic field vector and the flux rope axis) increasing from the center (axis) towards the periphery, and the structure typically expands as it travels outward from the Sun. Depending on the wrapping and inclination, the flux ropes can be classified into different types. There are two types of flux rope that exist, based on their inclination (i) low-inclined and (ii) high-inclined flux rope, and each type of flux rope can be further divided into four categories based on their polarities, i.e., rotation of field vector across the MC \cite{Burlaga1981,Bothmer1998,Mulligan1998,Nieves-Chinchilla2018}. A CME can drive an interplanetary shock if its speed in the solar wind frame exceeds the local magnetosonic speed. Sheath is a turbulent, compressed, and heated region of solar wind between the shock and the MC \cite{Tsurutani1988,Huttunen2002,Huttunen2004}. The properties of shock, sheath, and flux rope evolve as they propagate through the solar wind and therefore, studying them at different distances from the Sun, using multiple spacecraft, helps in understanding their evolution, influence on the local environment, and planetary space weather \cite{Temmer2014,Harrison2018,Amerstorfer2021,Xu2022,Chi2023,Chi2024}.

Although in situ observations provide several information about the dynamics and plasma properties of CMEs, such observations from a single vantage point are insufficient to infer the presence of all the substructures of a CME and its global properties \cite{Burlaga1981,Crooker1996,Nieves-Chinchilla2013}. The measured properties depend on the geometric selection effect--spacecraft trajectory through the nose or flank of the CME itself, impact parameter (the distance of closest approach to the flux rope axis), and flux rope orientation. The absence of any substructures in these observations does not strictly imply their absence within the CME itself, but rather may be a result of the spacecraft's unfavorable one-dimensional trajectory through the CME. For example, in situ observations, only 30\%-40\% CMEs exhibit flux rope (i.e., MC) substructure \cite{Gosling1990,Chi2016,Mishra2021a}, which is understood to be due to geometric selection effect \cite{Gopalswamy2006,Zhang2013}. The local single-point in situ observations of CMEs are also understood to provide inaccurate estimation of CME expansion speed \cite{Agarwal2024}. There are several studies focused on using multipoint in situ observations, with spacecraft radially aligned or longitudinally separated, to understand the 3D morphology, global structure, and temporal evolution of the CME substructures in the interplanetary (IP) medium \cite{Kilpua2011,Lugaz2018,Mishra2021,EmmaDavies2021a,Winslow2021,Mostl2022,Regnault2023,Regnault2024,Agarwal2025}. The accurate estimation of propagation and expansion dynamics of a CME in the pre-conditioned ambient medium due to another structure are crucial to connect the measurements with the thermodynamic evolution of CMEs in the IP medium \cite{Mishra2014,Mishra2015a,Winslow2016,Mishra2017,Kilpua2019,Mishra2018,Mishra2020,Khuntia2023,Khuntia2025}.

CMEs can expand as they propagate through the IP medium due to their higher internal pressure relative to the ambient solar wind \cite{Liu2006,Mishra2021a}. During propagation, CMEs may become distorted magnetized plasma structures, which can result in unequal forces acting on their LE and TE. Such force imbalances can cause the LE to propagate faster than the TE, thereby contributing, at least in part, to the radial size and expansion behaviour of CMEs \cite{Mishra2015}. A better understanding of CME expansion provides us insights into their thermal and kinetic properties \cite{Mishra2018,Mishra2020,Zhuang2023,Khuntia2023,Agarwal2024}. Additionally, CME expansion provides insight into the relative decrease in thermal and magnetic pressure within CMEs as they propagate through the heliosphere. The consequence of radial expansion of CMEs leads to an increase in their radial size, a decrease in magnetic field strength, and density of the CME \cite{Liu2005,Leitner2007,Gulisano2010,Good2015,Lugaz2020a,EmmaDavies2021}. Therefore, if the expansion speed of a CME constitutes a significant fraction of its propagation speed, the expansion can substantially influence its arrival time, radial size, geoeffective parameters, and the duration of associated geomagnetic perturbations \cite{Gulisano2010,Vourlidas2019,Mishra2021a,Agarwal2024}.

Generally, the temporal evolution of the MC speed observed by an in situ spacecraft exhibits a linearly decreasing profile, which is typically interpreted as evidence of CME expansion. However, this decrease can also be attributed to the deceleration of the CME. Moreover, the linear speed profile can be distorted in the rear part of the MC due to the presence of other solar wind structures, which can cause an apparent increase in the TE speed. To minimize the impact of such distortions on the expansion speed estimation, \citeA{Gulisano2010} adopted another approach where they used a linear function to fit the speed-time profile of the MCs without including the distorted region at the rear edge, and extrapolated this fit up to the rear boundary of the MC to obtain the propagation speed of the TE. The expansion speed is then determined using the LE and TE speeds derived from the fitted profile. The conventional method of estimating expansion speed, regardless of whether they assume a linear or non-linear speed-time profile, can have uncertainties in earlier studies. This is because the expansion speed, half the difference between the speeds of CME LE and TE, should be based on the simultaneous propagation speeds measurements of both substructures (LE and TE) \cite{Owens2005} while the arrivals of these substructures on a single in situ spacecraft are separated by several hours, with an average duration of approximately 20 hours at 1 AU \cite{Lepping2006}.

The conventional method and the method of \citeA{Gulisano2010} implies the assumption of a constant expansion speed (i.e., acceleration is neglected) during the passage of the CME over the in situ spacecraft \cite{Crooker1996,Owens2005,Demoulin2009,Richardson2010,Gopalswamy2015,Lugaz2017,Lugaz2020,Mishra2021a,Zhuang2023}. The conventional method of using single point in situ observations of CMEs cannot provide the CME instantaneous expansion speed at arrival instances of different CME substructure (LE, center, and TE) at in situ spacecraft. For example, a significant deceleration/acceleration in TE can lead to the overestimation/underestimation of expansion speed using the conventional method that assumes it is determined at the LE arrival time. These scenarios correspond to cases when the CME has a decreasing/increasing speed profile measured by a single point in situ spacecraft. Such an estimate of expansion speed is time-independent during the CME passage over the spacecraft. Further, the estimates of expansion speed from the conventional method have large errors when the CME exhibits a non-linear speed profile--the first half of the CME shows a decreasing speed profile while the second half shows an increasing trend, or vice versa.

Although the limitations of the conventional method for estimating CME time-independent expansion speed are acknowledged in the literature, it has been widely applied in both individual case studies and large-scale statistical analyses. For example, there are case studies on estimating radial expansion of CMEs from the conventional method, which are compared with the instantaneous expansion speed obtained from measuring the radial width of the CME from remote observations \cite{Savani2009,Savani2011,Savani2011a,Nieves-Chinchilla2012,Nieves-Chinchilla2013,Lugaz2020,EmmaDavies2021a}. Moreover, previous studies established the radial size and height relationship by estimating radial size from the CME’s propagation speed measured during its in situ passage. It is still uncertain how this relationship would be changed when the instantaneous expansion speed is used to estimate the CME’s radial size.

The statistical study of \citeA{Gopalswamy2015} found that the average expansion speed of CMEs from in situ observations in solar cycle (SC) 24 is around 50\% smaller than that in the previous cycle, possibly due to a decrease in the MC-to-ambient total pressure difference in SC24. The study of \citeA{Mishra2021a} found that the ratio of expansion to propagation speed in SC23 is $\sim$9\%, which decreased to $\sim$6\% in SC24. The expansion speed has also been used to explain some of the CMEs with smaller bulk speed driving the shock in the SC24 \cite{Lugaz2017}. These earlier studies have estimated the expansion speed using the conventional method from in situ observations, and the results derived may have some shortcomings due to not using the instantaneous expansion speed. This could also be a reason for the inconsistency in connecting the radial size and radial expansion speed of CMEs from remote observations with in situ observations \cite{Zhuang2023,Agarwal2024}.

Only a limited number of recent studies have explicitly addressed aging effects—namely, the influence of CME expansion, acceleration, distortion, and deflection during a spacecraft’s passage through the structure. Because of these effects, in situ observations represent a one-dimensional time series sampled along the spacecraft trajectory across the CME’s radial extent, rather than a snapshot of the CME at a single instant. Consequently, measurements taken sequentially from the CME’s front to its rear correspond to progressively evolved states of different substructures, rather than at the same evolutionary stages at a single instant. As a result, direct comparisons among different CME substructures using 1D in situ data are inherently non-instantaneous and must be interpreted with caution \cite{Demoulin2020,Regnault2023,Regnault2024a,Regnault2024,Agarwal2024}. Particularly, the work of \citeA{Regnault2024} and \citeA{Agarwal2024} focused on estimating the time-dependent instantaneous expansion speed of CMEs. The instantaneous expansion speed of a CME can be estimated by utilizing the radially aligned multipoint in situ observations, provided the propagation speeds of any two substructures are measured simultaneously \cite{Regnault2024}.

Moreover, \citeA{Agarwal2024} proposed a non-conventional method to estimate the instantaneous expansion speed by utilizing the propagation speed of any two substructures (LE, center, and TE) at the same instance, even by using a single point in situ spacecraft. The method also estimates the radial size of the CME and the distance traveled by any substructure during any two instances. This method assumes a constant acceleration of any CME substructure during the passage of the CME at in situ spacecraft. For the non-conventional method, we coin the name Constant Acceleration Accounted Perspective (CAAP) for estimating the time-dependent instantaneous expansion speed at any time during the passage of the CME at a single-point in situ spacecraft. Our method is different from the conventional method, which neglects the acceleration and estimates time-independent expansion speed during the CME passage at the spacecraft. Although, \citeA{Demoulin2020} highlighted that the in situ measured propagation speed profile across a MC at a single spacecraft reflects the combined effects of the acceleration/deceleration of the CME center and the expansion of the CME during the spacecraft crossing. However, to estimate the expansion factor, their method assumes a constant speed of the MC center (i.e., the absence of bulk acceleration) throughout the passage. Therefore, the method employed in our study differs fundamentally from \citeA{Demoulin2020}. The proposed non-conventional method (CAAP) is relatively new in the literature and has not been widely implemented in several case studies.

In the direction of strengthening the reliability of the CAAP method, it is crucial that we test the method using multipoint in situ observations of any two substructures simultaneously. Since such a validation of the method has not yet been done in the literature, we focus on this goal in the present study. For this purpose, we have selected a CME of 3-5 November 2021 observed from radially aligned in situ spacecraft Solar Orbiter (SolO) at 0.85 AU and Wind at 0.98 AU. Both spacecraft are separated by 0.13 AU radially; the angular separation between the spacecraft is 2.3$^\circ$ (1$^\circ$ in longitude and 2.1$^\circ$ in latitude on 4 November 2021). Although this small angular separation implies that each spacecraft may sample slightly different portions of the CME, it remains well within the theoretical spatial coherence limit for large-scale CME structures \cite{Owens2017}. However, radial evolution of the CME can still influence its coherent structure. Therefore, differences in MC characteristics between the two spacecraft are primarily attributed to the radial evolution of the CME. Furthermore, \citeA{Rossi2025} demonstrated that the discrepancies in plasma properties observed at the two locations can be explained largely by considering radial evolution and cannot be explained solely by considering the different sampled portions of the CME. Thus, we assume that both spacecraft largely sampled the same CME features. The selected CME provides an opportunity to validate the efficacy of the CAAP method for estimating instantaneous expansion speed. This CME, classified as an MC, is a unique and rare event because its centre and TE are simultaneously observed by Wind and SolO, respectively. Such simultaneous observations of the center and TE help to directly estimate the time-dependent instantaneous expansion speed of the MC at a particular instance. It also offers an opportunity to assess whether in situ observations made prior to the L1 point are sufficient for reliable prediction of magnetic and plasma parameters at the L1 point.

\begin{figure}
    \centering
    \includegraphics[scale= 0.21,trim={0cm 0cm 0cm 0cm},clip]{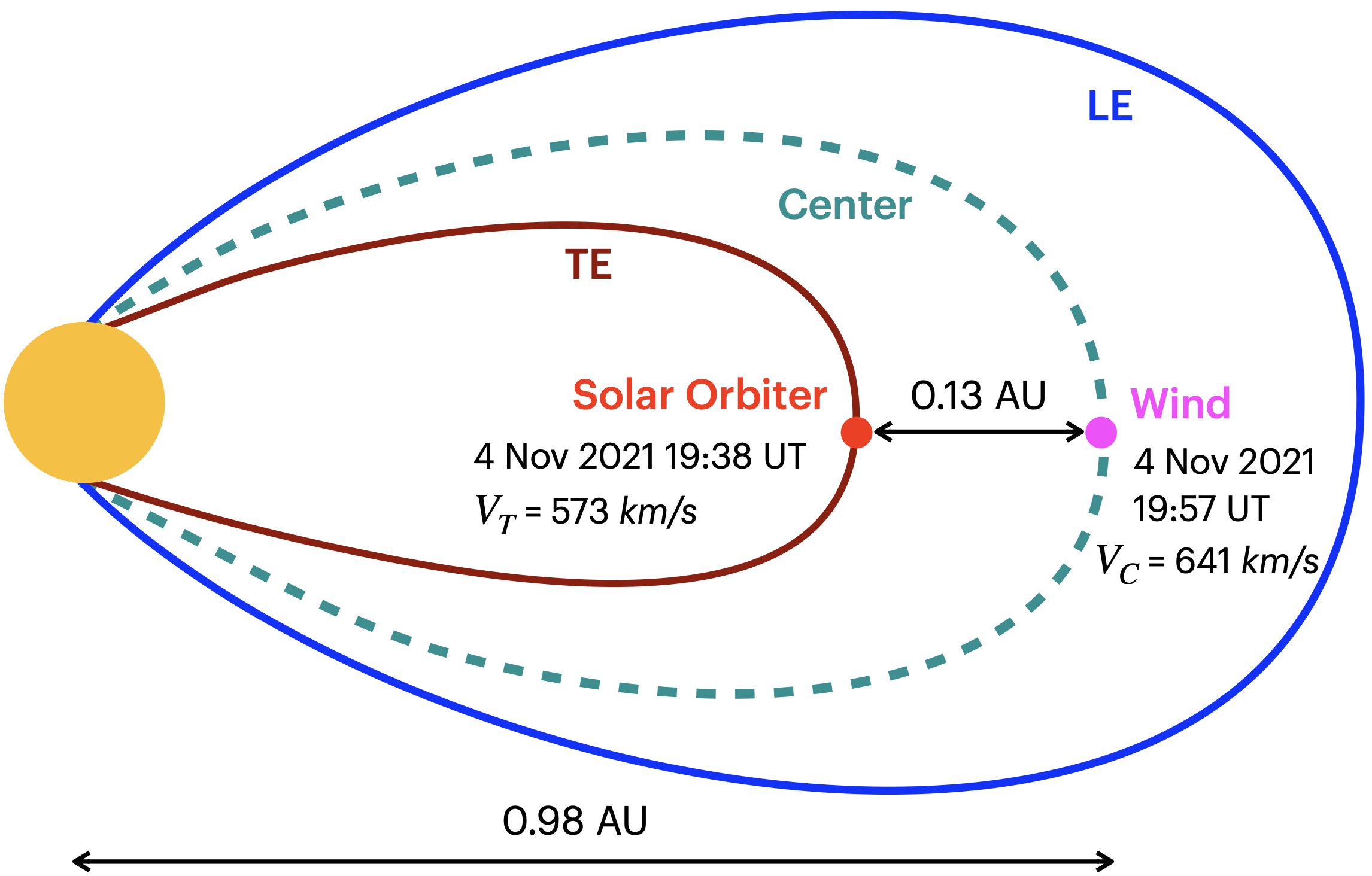}
    \caption{The figure depicts the simultaneous observation of the different substructures, center and TE, by Wind and SolO, respectively. The blue, green, and maroon represent the MC substructures LE, center, and TE. The yellow circle represents the Sun. The red and magenta circle shows the radial location of the SolO and Wind, respectively, from the Sun in the IP medium.}
    \label{fig:space_posi}
\end{figure}

Figure~\ref{fig:space_posi} depicts the rare scenario when two substructures of the CME are measured simultaneously by radially aligned in situ spacecraft. The MC associated with the selected CME has recently been studied \cite{Regnault2024,Rossi2025}. The study of \citeA{Regnault2024} estimated the instantaneous expansion speed of the MC using simultaneous in situ measurements from SolO and Wind and examined its evolution between the two spacecraft, reporting an unusual acceleration of the MC during its propagation from SolO to Wind. To identify possible external forces responsible for the observed acceleration, \citeA{Rossi2025} applied and tested the Extended Drag-Based Model on this CME. However, it is important to compare the estimates of MC's instantaneous expansion speed in \citeA{Regnault2024} to that of expansion speed estimates from a non-conventional CAAP method \cite{Agarwal2024}, which can account for acceleration of CME substructures and derive their propagation speeds throughout its passage from one distance to another. Such a comparison of estimates obtained directly from the simultaneous in situ measurements and those derived from our CAAP can help examine the efficacy of the CAAP method. Additionally, our study also examines the evolution of the MC axis orientation or its magnetic flux from SolO to Wind.

Although there are earlier studies discussing the shock associated with the selected MC \cite{Trotta2023,Regnault2024}, a comprehensive analysis of the shock and its characteristics at both spacecraft still remains to be reported. The study of \citeA{Trotta2023} primarily investigated the formation of the shocklets in the upstream solar wind observed at Wind, which were not present at SolO, while the study of \citeA{Regnault2024} discusses the actual and expected arrival of the shock at the two spacecraft. In this context, we provide a comprehensive analysis of the shock in terms of its speed, geometry, and Mach number to understand the unusual evolutionary nature of the shock between SolO and Wind. A detailed description of the multipoint in situ observations, the validation of the CAAP method for estimating the instantaneous expansion speed using single-point observations, and the evolution of CME substructures are described from Sections~\ref{sec:obs} to \ref{sec:evocmesub}. Finally, Section~\ref{sec:resdis} outlines our results and discussions.

\section{Observations and Analysis Methodology} \label{sec:obs}

\begin{figure*}
    \centering
    \includegraphics[scale= 0.295,trim={0cm 0cm 0cm 0cm},clip]{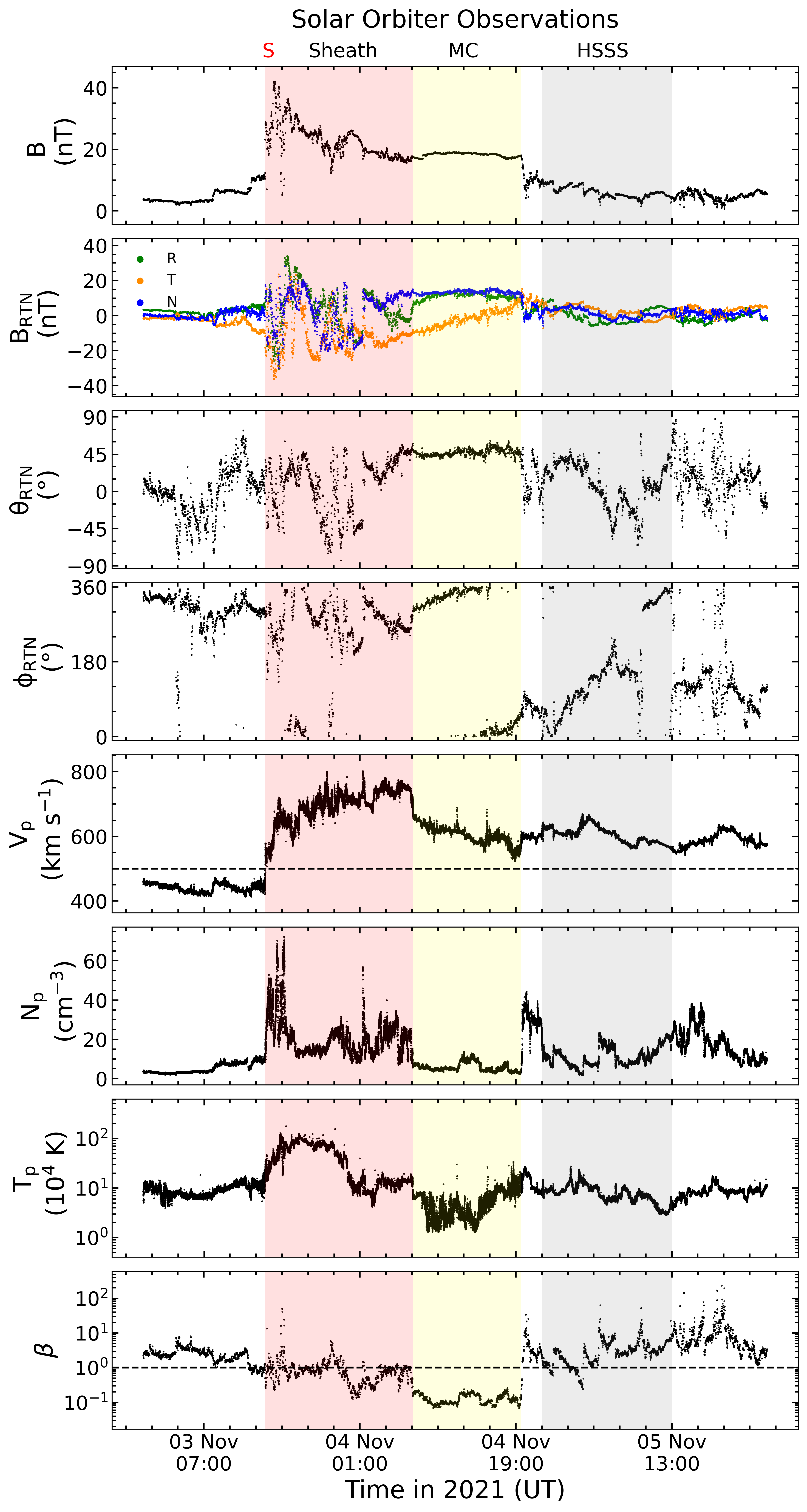}
    \includegraphics[scale= 0.295,trim={0cm 0cm 0cm 0cm},clip]{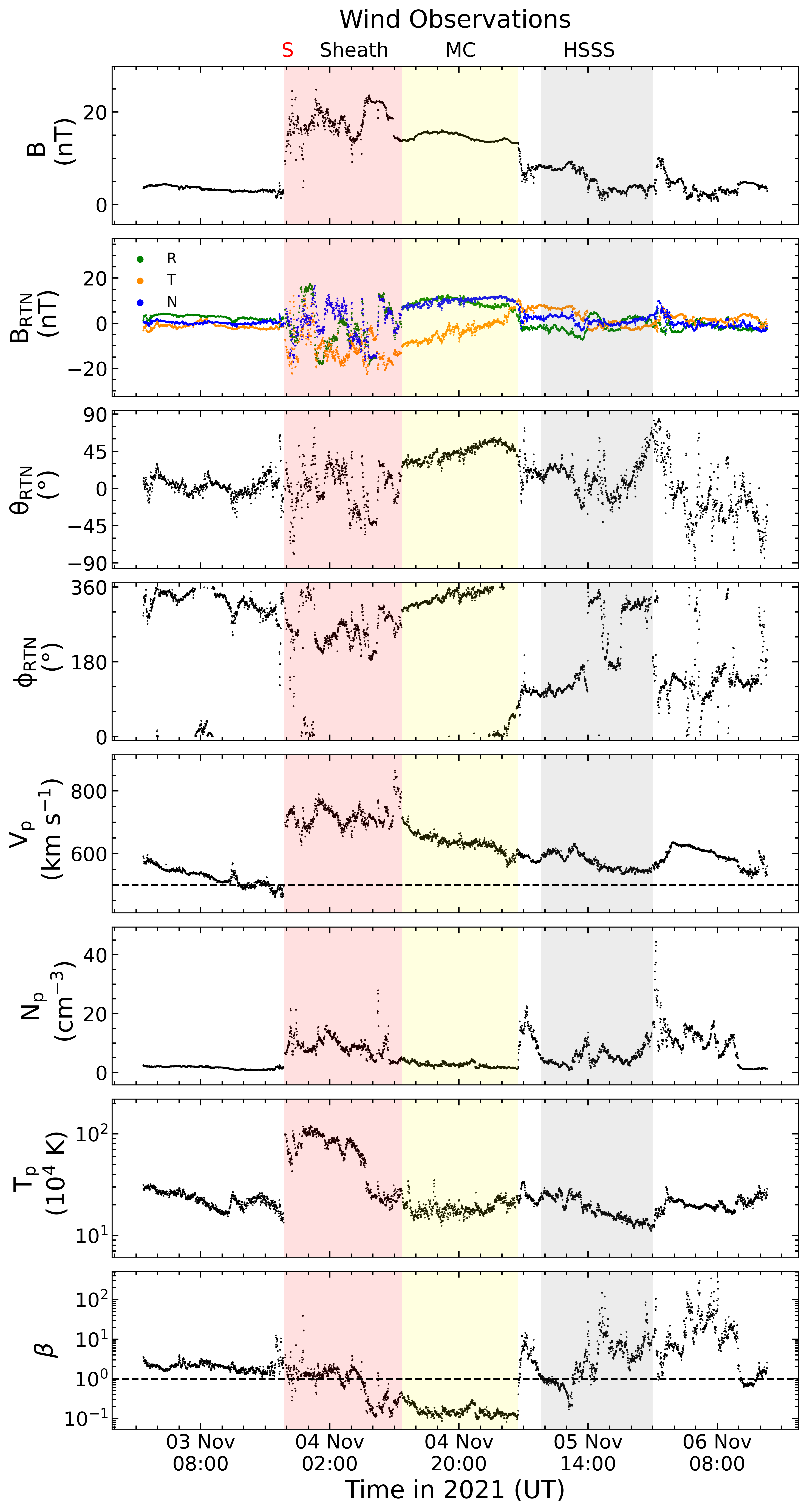}
    \caption{The top-to-bottom panels depict the variations in the total magnetic field, the components of the magnetic field vector in the RTN coordinate system, latitude and longitude of the total magnetic field vector, proton speed, proton density, proton temperature, and plasma beta, respectively, at SolO and Wind in the left and right columns, respectively. Transparent fill areas with red and yellow represent the sheath and MC duration during the passage of the CME on both spacecraft.}
    \label{fig:insitu_solo_wind}
\end{figure*}

We focus on validating the non-conventional method (CAAP) applied to single-spacecraft observations by using the observations from two radially separated spacecraft. The validation involves finding the time-dependent instantaneous expansion speed of the MC using propagation speed estimates of its substructures obtained through our CAAP method, and comparing these with the measured instantaneous expansion speed estimated from simultaneous in situ measurements of the MC's center by Wind and its TE by SolO. We also focus on examining the differences in the properties of CMEs' substructures measured at SolO and Wind. The in situ observations of the CME from SolO and Wind are shown in the left and right panels of Figure~\ref{fig:insitu_solo_wind}, respectively. The top-to-bottom panels show the total magnetic field, components of the magnetic field vector in the RTN coordinate system, latitude ($\theta$), and longitude ($\phi$) of the total magnetic field vector, speed, density, temperature, and plasma beta. The transparent red and yellow region depicts the sheath and MC, respectively. The criteria used for identifying these CME substructures are discussed in Section~\ref{sec:insituobs}. For the magnetic field and plasma measurements of the CME at SolO, we have utilized the MAG and SWA instruments \cite{Horbury2020,Owen2020}. For the magnetic field and plasma parameters except for temperature at Wind, we have utilized the MFI and SWE instruments, while for temperature, the 3DP instrument is utilized \cite{Lepping1995,Ogilvie1995,Lin1995} as the temperature from the SWE, in comparison to the 3DP instrument, is unusually high in the MC region \cite{Regnault2024}. 

\subsection{In Situ Observations of the CME at Solar Orbiter and Wind} \label{sec:insituobs}

\begin{table}
\caption{The top and bottom panels of the table list the arrival time and radial size of CME substructures at SolO and Wind, along with the differences between them.}
\centering
\begin{tabular}{cccc}
\hline
\multicolumn{4}{c}{In Situ Arrival Time of CME Substructures} \\
\hline
Substructure & SolO ($t$) [UT] & Wind ($t'$) [UT] & $\Delta t = t' -t$ [hr]\\ 
\hline
Shock & 3 Nov 14:03:30 & 3 Nov 19:34:23 & 5.51 \\
LE & 4 Nov 07:09 ($t_1$) & 4 Nov 12:06 ($t_1'$) & 4.97 \\
Center & 4 Nov 13:09 ($t_2$) & 4 Nov 19:57 ($t_2'$) & 6.81 \\
TE & 4 Nov 19:38 ($t_3$) & 5 Nov 04:14 ($t_3'$) & 8.6 \\
\hline
     \multicolumn{4}{c}{Radial Size of CME substructures [$R_\odot$]} \\
\hline
Substructure & SolO ($R$) & Wind ($R'$) & $\Delta R = R' - R$\\ 
\hline
MC & 39.3 & 53.5 & 14.2 \\
Sheath & 61.9 & 61.3 & -0.6 \\
\hline 
\end{tabular}
\label{tab:arr_rad_tab}
\end{table}

\begin{figure}
    \centering
    \includegraphics[scale = 0.6,trim={0cm 0cm 0cm 0cm},clip]{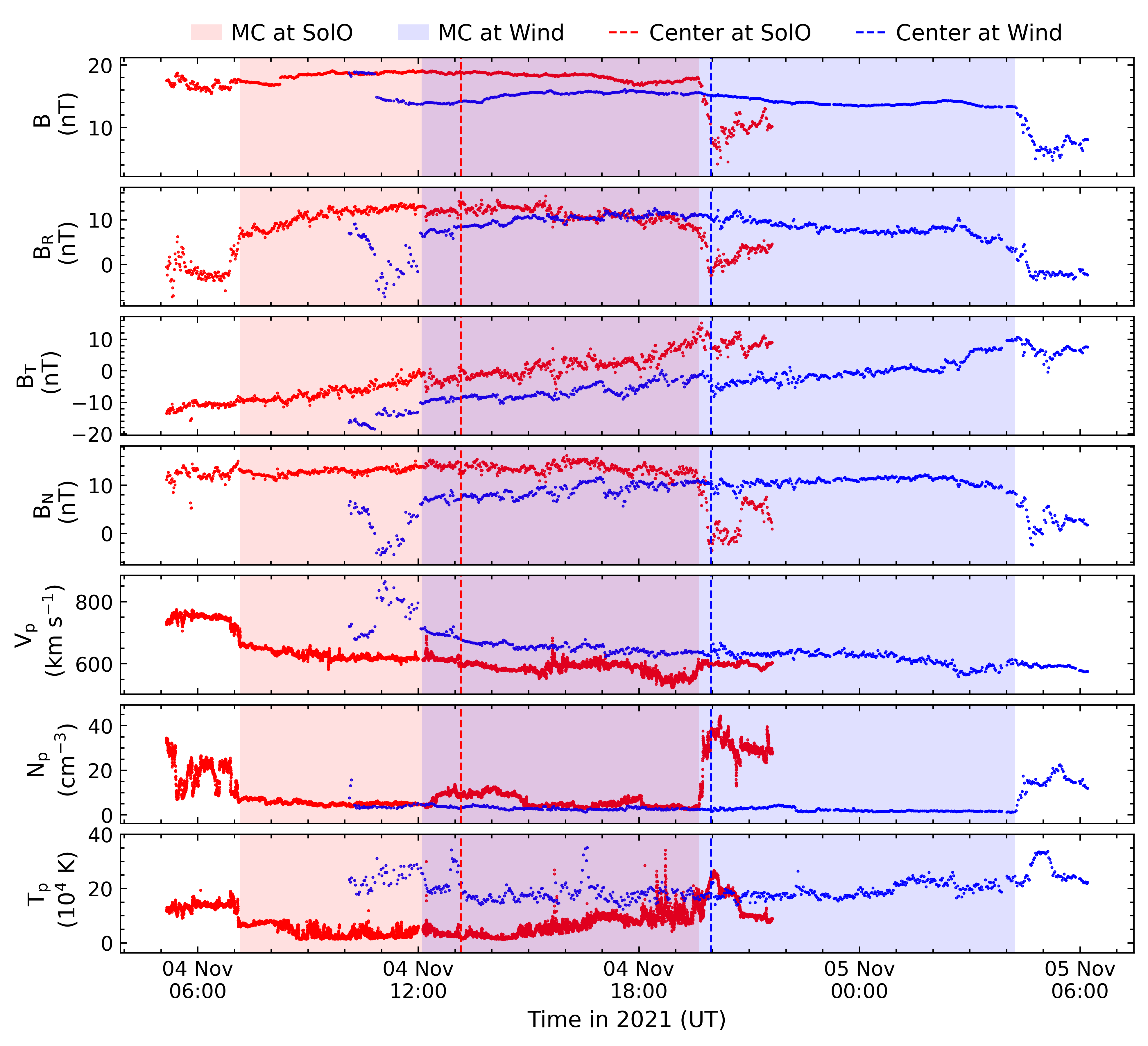}
    \caption{The top-to-bottom panels depict the magnetic field and plasma parameters, which are primarily used to identify the MC region at both spacecraft. The schematic overlays the MC intervals at SolO and Wind in red and blue shaded regions, respectively. The red and blue dashed lines represent the arrival of the center at SolO and Wind, respectively.}
    \label{fig:overplot}
\end{figure}

To identify the boundaries of CME substructures, we scrutinize the in situ data by employing multiple signatures simultaneously \cite{Zurbuchen2006} and cross-check the determined boundaries with those reported by \citeA{Regnault2024}. The boundaries identified in our analysis are consistent with the boundaries presented in their study. The arrival times of the shock at SolO and Wind are listed in the top panel of Table~\ref{tab:arr_rad_tab}, which also includes the identified start (LE arrival) and end (TE arrival) boundaries of the MC at both spacecraft. The MC boundaries are identified by scrutinizing the magnetic field and plasma parameters that fulfill the conditions of an MC, particularly focused on the smooth rotation of the magnetic field vectors ($B_{RTN}$). Figure~\ref{fig:overplot} displays the magnetic field and plasma parameters (speed, density, and temperature) of the MC region, extending two hours before and after the MC region at both spacecraft. The MC at SolO and Wind is illustrated in red and blue shaded regions, respectively. We notice that the depicted magnetic field and plasma parameters show similar trends at Wind as observed at SolO, with differences in their magnitude due to the temporal evolution of the MC. This implies that the same substructures (LE, center, and TE) are identified at both spacecraft.

However, the identified boundaries of MC at Wind are different from the start and end boundaries of MC in Richardson \& Cane Catalog (\url{https://izw1.caltech.edu/ACE/ASC/DATA/level3/icmetable2.htm}) and DREAMS Catalog (\url{https://space.ustc.edu.cn/dreams/wind_icmes/index.php}). This discrepancy arises because we determine the boundaries based strictly on the rotation of the magnetic field vectors, represented by $\theta_{RTN}$ and $\phi_{RTN}$ in the third and fourth panels of Figure~\ref{fig:insitu_solo_wind}. Further, we have identified and marked the arrival of the MC center as the size center (that divides the radial size of the MC at the arrival of LE into two equal parts) of the MC at both spacecraft \cite{Agarwal2024}. The red and blue dashed lines in Figure~\ref{fig:overplot} depict the arrival of the MC center at SolO and Wind, respectively. The radial size of CME substructures is calculated by utilizing the trapezoidal rule of integration on the in situ measured speed-time profile as follows:

\begin{equation} \label{equ:radialsize}
   R = \int_{t_0}^{t_n} V \, dt  \approx \sum_{i =0}^{n-1} \left(\frac{V_i + V_{i+1}}{2}\right)\, (t_{i+1} - t_i)
\end{equation}

where $R$ represents the radial size of the substructure, $V_i$ represents the speed of the substructure at time $t_i$, i = 0, 1, 2,......, n-1, during its passage over the spacecraft. The estimated radial size of CME substructures (sheath and MC) is listed in the bottom panel of Table~\ref{tab:arr_rad_tab}. The identified boundary of the MC center at SolO and Wind is 4 November 2021 at 13:09 UT and 4 November 2021 at 19:57 UT, respectively. The radial size of the MC estimated using the trapezoidal rule reflects the size at the arrival of the MC's LE at the in situ spacecraft. Although the size of an expanding MC continuously increases during its passage over the spacecraft, its propagation speed also varies across its complete structure, which is higher at initial times and monotonically smaller at later times across the CME passing duration. Such expansion-induced changes in CME size will appear in the propagation speeds measured in situ, causing the CME center to reach the spacecraft earlier relative to its time-center (the midpoint of the CME’s temporal profile) than it would be expected for a non-expanding CME. The changes in the propagation speed of substructures (from the front to back) influence the total time taken by the MC during its passage over the spacecraft. It is evident that our estimate of the MC radial size is at the arrival time of the MC's LE at the spacecraft.

The arrival of different substructures (shock, sheath, and MC) at SolO and Wind indicates changes in their speed while traveling between the two spacecraft. Also, there is a noticeable change in their size at SolO and Wind, which requires explanations considering the physical processes the CME is undergoing. The discussion on these aspects of the evolution of CME substructures between SolO and Wind will be done in Section~\ref{sec:evocmesub}. In the following section, we first focus on estimating the instantaneous expansion speed of the selected MC using simultaneous in situ observations. We will also estimate the instantaneous expansion speed at both spacecraft using the CAAP method and compare our estimates with the measured value to validate the efficacy of the CAAP method.

\subsection{Instantaneous Expansion Speed: In Situ Observations and Non-conventional Method} \label{sec:instexpa}

From a space weather perspective, the expansion speed of an MC is a critical parameter, as it governs both its radial extent and the duration of associated geomagnetic disturbances. The instantaneous expansion speed of the selected MC is estimated by leveraging its unique observations at the multiple spacecraft: the arrival of the TE at SolO ($t_3$) is nearly simultaneous with the arrival of the center at Wind ($t_2'$), i.e., $t_2' \sim t_3$, as outlined in the top panel of Table~\ref{tab:arr_rad_tab}. Although the arrival of the center and TE of the CME at Wind and SolO, respectively, are not exactly simultaneous--with a time difference of nearly 19 minutes--the resulting instantaneous expansion speed remains a valid approximation in comparison to the expansion speed from the conventional method. Therefore, we can compute its approximate value of instantaneous expansion speed ($V_{\mathrm{exp_{inst}}}$) by using the nearly simultaneously measured speed of MC center ($V_{{C_{Wind}}}$) and TE ($V_{{T_{SolO}}}$) as follows:

\begin{equation} \label{equ:instan_exp}
    V_{\mathrm{exp_{inst}}}(t_2' \sim t_3) = V_{{C_{Wind}}} (t_2') - V_{{T_{SolO}}} (t_3)
\end{equation}

To estimate the average speeds of the CME substructures (LE, center, and TE), we account for inherent fluctuations in the in situ speed measurements as the CME passes over the spacecraft. Rather than relying on speed values at their sharp arrival times, we adopt a time-averaged approach over a defined duration, assuming a finite thickness for each substructure. Recent studies by \citeA{Temmer2022,Agarwal2024} have suggested that CME substructures do not exhibit sharp boundaries but instead possess a gradual transition with measurable thickness. In line with these findings, we define a duration of 10 minutes for each substructure--the LE, center, and TE--to compute their average speeds based on in situ observations. The computed average speed over this duration at both spacecraft is shown in the bold font in the top panel of Table~\ref{tab:speed_tab}. Utilizing Equation~\ref{equ:instan_exp} and the measured speed of the center and TE at Wind and SolO, respectively, the estimated instantaneous expansion speed of the MC is 68 $km~s^{-1}$.

We aim to compare the instantaneous expansion speed obtained directly from the simultaneously in situ observed propagation speeds of the center and TE at multi-spacecraft (Wind and SolO) with the instantaneous expansion speed derived from our CAAP method. We emphasize that our CAAP method uses the observed propagation speed exclusively from a single spacecraft. Since the measurements of individual substructures (LE, center, and TE) by a single-point in situ spacecraft are obtained at different instances, the CAAP method can estimate the propagation speeds of each substructure at the same instance. This requires the acceleration of the individual substructures as an external input parameter. The acceleration of the individual substructures could be derived from single spacecraft in situ observations assuming a certain thickness of the substructures, as adopted in \citeA{Agarwal2024}. However, such acceleration estimates (from the single-point in situ observations) can introduce uncertainties in the CAAP method if there is a presence of other solar wind structures at its front and/or rear edges. In such a situation where a CME is clearly experiencing a drag force due to its differing propagation speed from the background solar wind medium, one can also use drag-based model (DBM) to estimate the acceleration of CME to use as an input in the CAAP method. Since our goal is to validate the CAAP method, we prefer not to rely on acceleration estimates derived from the DBM or from the method adopted in \citeA{Agarwal2024}, as these approaches may introduce additional uncertainties. Instead, we utilize the availability of multi-spacecraft observations (SolO and Wind), which provide speed measurements of the same substructure at two locations, to estimate the mean acceleration of the substructures. These estimates are then used as inputs to the CAAP method to obtain the instantaneous expansion speed.

\setlength{\cmidrulewidth}{0.5pt}
\begin{table}
\caption{The top panel of the table lists the propagation speed of CME substructures at SolO and Wind, where values directly measured from in situ observations are in bold, and those estimated using our non-conventional method are in normal font. The bottom panel of the table lists the expressions used to estimate the instantaneous expansion speed, estimated instantaneous expansion speed at SolO and Wind in the first, second, and third columns, respectively.}
\centering
\begin{tabular}{ccccccc}
\hline
\multicolumn{7}{c}{Speed of CME Substructures at Different Instances [$km~s^{-1}$]} \\
\hline
{\multirow{2}{*}{Substructure}} & \multicolumn{3}{c}{SolO} & \multicolumn{3}{c}{Wind}  \\
\cmidrule(lr){2-4}
\cmidrule(lr){5-7}
& ${t_1}$ & ${t_2}$ & ${t_3}$ & ${t_1'}$ & ${t_2'}$ & ${t_3'}$ \\
    \hline
    {LE} & \textbf{661} & 718 & 780 & \textbf{708} & 782 & 860 \\
    {Center} & 594 & \textbf{616} & 640 & 612 & \textbf{641} & 671 \\
    {TE} & 522 & 546 & \textbf{573} & 542 & 574 & \textbf{608} \\
    \hline
    \multicolumn{7}{c}{Expansion Speed: Non-conventional and Conventional Method [$km~s^{-1}$]} \\ \hline
    \multicolumn{3}{c}{\multirow{3}{*}{Expression using}} & \multicolumn{2}{c}{Inst. expansion speed:} & \multicolumn{2}{c}{Expansion speed:} \\
    \multicolumn{3}{c}{\multirow{3}{*}{speed of substructures}} & \multicolumn{2}{c}{Non-conventional method} & \multicolumn{2}{c}{Conventional method} \\
    \cmidrule(lr){4-5}
    \cmidrule(lr){6-7}
    & & & {TE arrival} & {Center arrival} & \multirow{2}{*}{SolO} & \multirow{2}{*}{Wind}\\ 
    & & & {at SolO ($t_3$)} & {at Wind ($t_2' \sim t_3$)} & &\\
    \hline
    \multicolumn{3}{c}{$V_L-V_C$} & {140}  & {141} & 45 & 67\\
        \multicolumn{3}{c}{$V_C - V_T$} & {67} & {67} & 43 & 33\\
    \multicolumn{3}{c}{$\frac{V_L - V_T}{2}$} & {103} & {104} & 44 & 50\\
\hline
\end{tabular}
\label{tab:speed_tab}
\end{table}

\begin{figure}
    \centering
    \includegraphics[scale = 0.32,trim={0.2cm 0cm 0cm 0cm},clip]{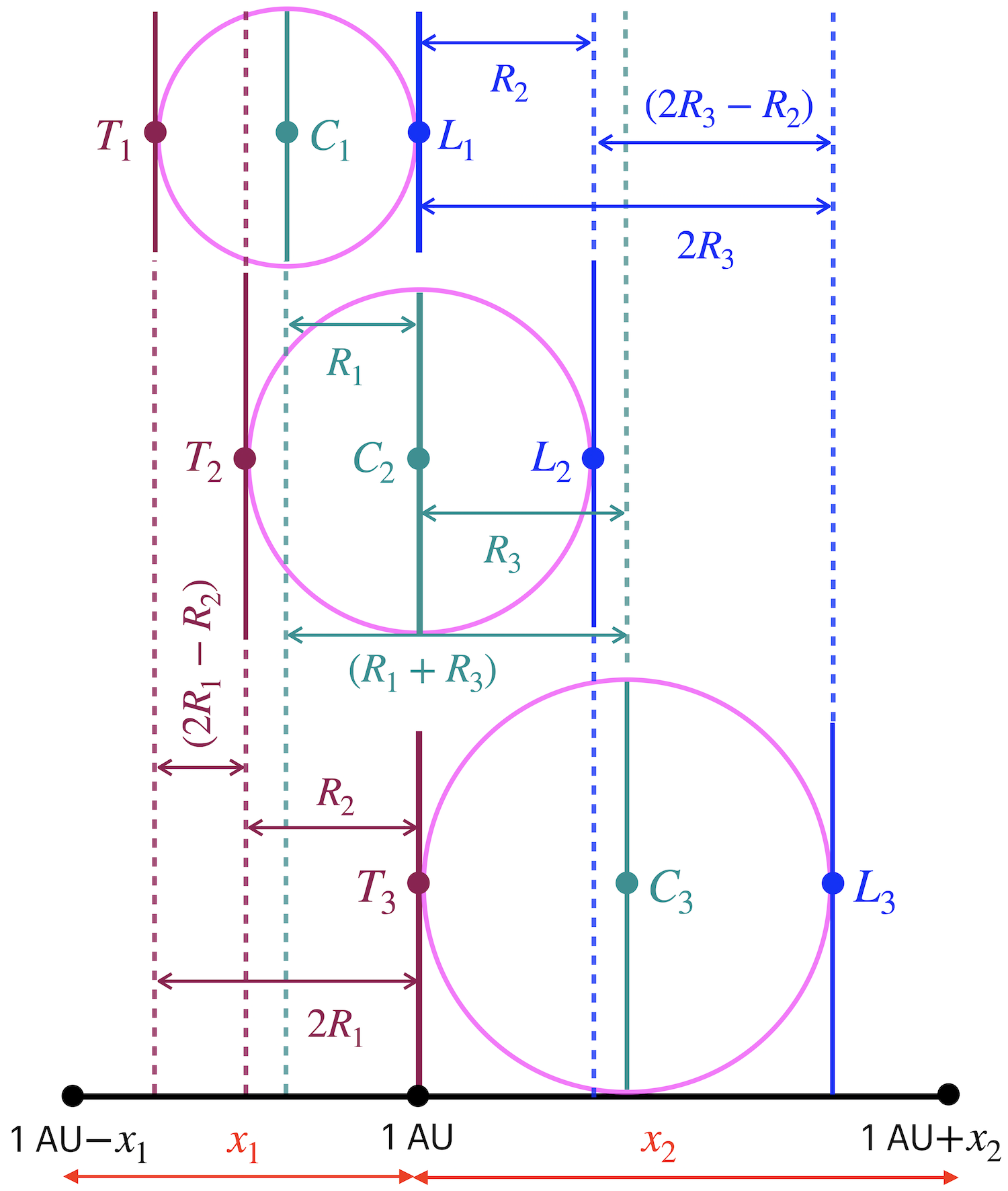}
    \caption{The schematic depicts the evolution of an expanding MC during its passage over the in situ spacecraft. The magenta circles represent the geometry of an MC in the plane of an in situ spacecraft. The blue, green, and maroon vertical lines denote the LE, size center, and TE of the MC, respectively. In the schematic, the in situ spacecraft is positioned at 1 AU and marked with a filled circle on the horizontal black line with two additional distances, one greater than 1 AU (1 AU + $x_2$) and one less than 1 AU (1 AU - $x_1$). The top, middle, and bottom panels represent the arrival of the LE (L), center (C), and TE (T), respectively, at the in situ spacecraft at different instances. The left-right arrow represents the distance traveled by each CME substructure during any two instances (adapted from \citeA{Agarwal2024}).}
    \label{fig:non-conven_fig}
\end{figure}

\begin{figure}
    \centering
    \includegraphics[scale = 0.1,trim={0cm 0cm 0cm 0cm},clip]{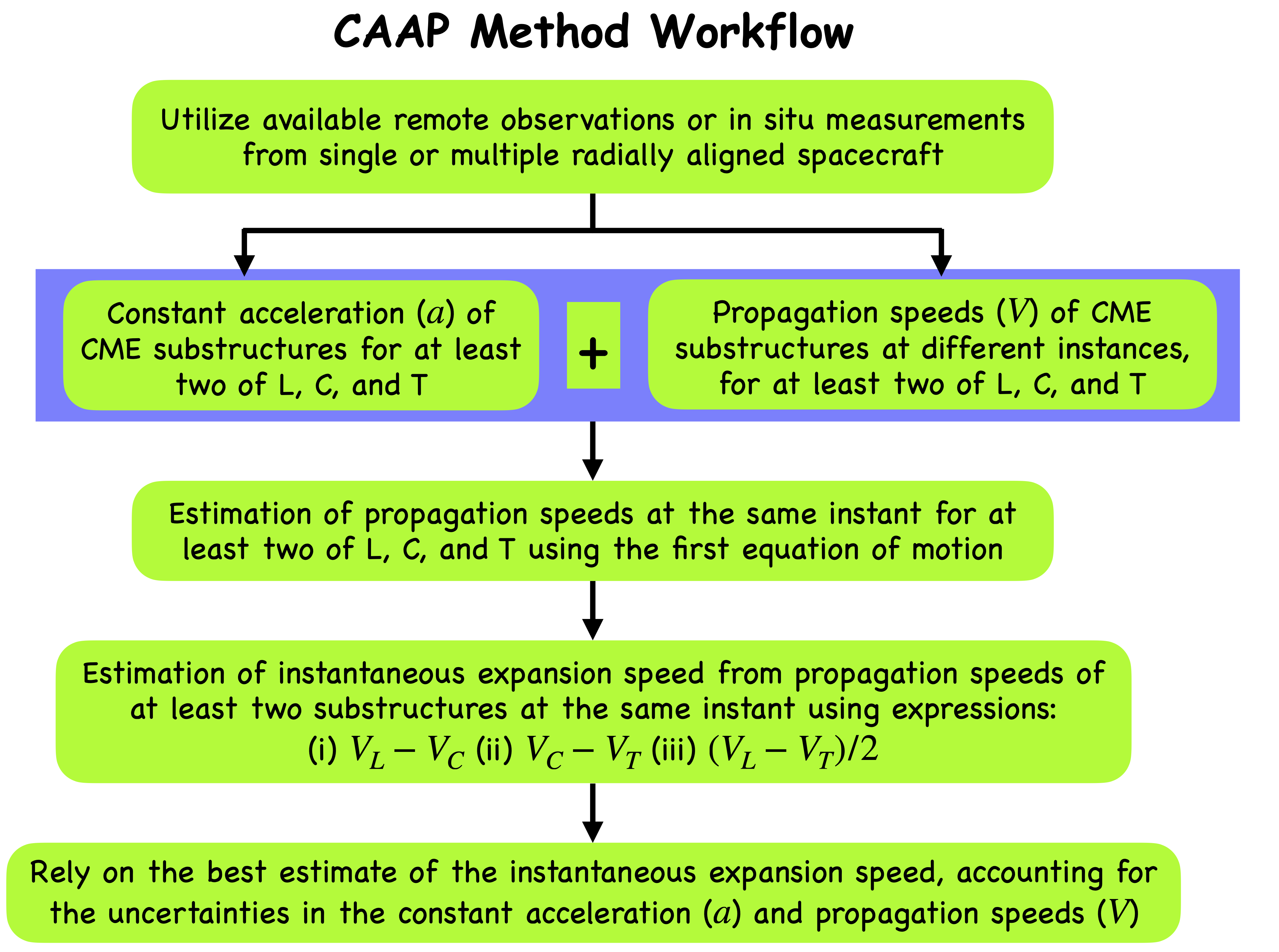}
    \caption{The schematic depicts the workflow of the CAAP method for estimating the instantaneous expansion speed ($V_{inst\_exp}$) of a CME using the constant acceleration and measured propagation speeds of its substructures. The symbols L, C, and T denote the CME substructures LE, center, and TE, respectively.}
    \label{fig:CAAP_workflow}
\end{figure}

We implement the non-conventional method on the single-point in situ observations of both the SolO and Wind spacecraft. Figure~\ref{fig:non-conven_fig} illustrates the pictorial representation of an expanding MC during its passage over the in situ spacecraft, which is assumed to be located at 1 AU. The magenta color circle represents the theoretical geometry of MC in the plane of the in situ spacecraft. The horizontal black line marks the position of the in situ spacecraft at 1 AU with two additional distances, one greater than 1 AU (1 AU + $x_2$) and one less than 1 AU (1 AU - $x_1$). The blue, green, and maroon vertical lines represent the LE, center, and TE of the MC. The top-to-bottom panels represent the arrival of LE (L), center (C), and TE (T) on the in situ spacecraft (located at 1 AU) at three different instances $t_1$, $t_2$, and $t_3$, respectively. The radii of the expanding MC at three different instances are $R_1$, $R_2$, and $R_3$. The left-right arrow denotes the distance traveled by any substructure during any two instances. Figure~\ref{fig:CAAP_workflow} illustrates the workflow of the CAAP method for estimating the instantaneous expansion speed of the CME. The calculations presented below adhere to this workflow. The CAAP method assumes constant acceleration of CME substructures to estimate their speed at different instances using the first equation of motion as follows:

\begin{equation}\label{equ:speed}
    {V_{F}({t_j}) = V_{F}({t_i}) + a_{F}{t_{ji}}}
\end{equation}

where the subscript $F$ denotes CME substructures (any of $L$, $C$, and $T$) and $t_{ji}$ is the duration between two instances as $t_j - t_i$ (for $j > i$). The symbols $V$ and $a$ denote the speed and constant acceleration of substructures. The acceleration of each substructure is estimated from the two-point measurements of propagation speed at SolO and Wind. The estimated acceleration represents the mean acceleration experienced by the CME substructures during their propagation from SolO to Wind. The acceleration is assumed to be constant during the passage of the CME at a single spacecraft and also during its journey from the SolO to Wind. The acceleration is computed as follows;

\begin{equation}
    a_F = \frac{(V_F)_{Wind} - (V_F)_{SolO}}{(t_F)_{Wind} - (t_F)_{SolO}}
\end{equation}

The computed accelerations of LE, center, and TE are 2.63 $m~s^{-2}$, 1.02 $m~s^{-2}$, and 1.13 $m~s^{-2}$, respectively. From these, it is evident that all CME substructures undergo acceleration during their transit from SolO to Wind, consistent with the findings of previous studies \cite{Regnault2024,Rossi2025}. The positive acceleration of fast-speed CME substructures is uncommon at distances that are farther from the Sun \cite{Zhang2006}. However, the selected MC appears to be accelerated and compressed at SolO, likely due to the presence of the high-speed solar wind stream (HSSS) at its rear edge \cite{Rossi2025}. Moreover, the estimated acceleration of the LE is higher than that of the TE, even though the MC is subjected to a high-speed solar wind at its rear. This behavior may result from the injection of additional magnetic flux into the rear of the MC between SolO and Wind (see Section~\ref{sec:flux_radius}), which can delay the arrival of the TE at Wind. Consequently, the TE is experiencing the drag force for a longer duration than the LE, leading to a lower mean acceleration of the TE compared to the LE.

We note that the study by \citeA{Regnault2024} reported no strong evidence of an HSSS or compression at the rear of the MC. However, the presence of an HSSS at the rear of the MC is suggested by typical observational indicators, including unusually high solar wind speed compared to the surrounding background ($\sim$500 $km~s^{-1}$), increased density and temperature, and elevated plasma beta \cite{Grandin2019,Remeshan2026}. A more definitive confirmation would have been possible if the heavy-ion charge state ratios were found to be lower than those in the slow solar wind \cite{Borovsky2016}; however, such measurements in the HSSS region are not available for this event at either spacecraft. In Figure~\ref{fig:insitu_solo_wind}, we have marked the plausible region of HSSS with a transparent gray area at both spacecraft. Additionally, using remote EUV observations, we note that around the end of October, a coronal hole was extended at approximately 40$^\circ$ latitude and 0$^\circ$ longitude, which could be a potential source of the high-speed stream. A detailed analysis of the remote observations of the coronal hole and its wind stream reaching to SolO and Wind is beyond the scope of the present study. Furthermore, from the study of \citeA{Li2022}, it is confirmed that during this interval, the solar wind is not pristine.

We also note that the solar wind speed ahead of the CME at SolO is slower by approximately 180 $km~s^{-1}$ than the speed at the MC center, while the solar wind behind the MC is slower by about 15 $km~s^{-1}$. We note that based on the current understanding of CME-solar wind interaction \cite{Vršnak2013}, such conditions (despite the presence of HSSS around the CME), would favor CME deceleration rather than the observed acceleration. Similarly, at Wind, the solar wind speed behind the MC is slower by about 70 $km~s^{-1}$ relative to the MC center speed, while the upstream solar wind is slower by approximately 115 $km~s^{-1}$. Under these circumstances, the observed acceleration of CME substructures from SolO to Wind cannot be readily explained by conventional drag-based models of CME propagation. The discussion suggests that accurately estimating the acceleration of a CME from single-point in situ observations can be difficult for some events, and using such estimates as input to the CAAP method may introduce uncertainties in its results. This highlights the sensitivity of the CAAP method and emphasizes that its successful application depends on the reliability of the input acceleration.

In a recent study, \citeA{Rossi2025} addressed this observed acceleration of the CME by introducing an additional acceleration term in the DBM to represent residual CME acceleration arising from factors such as magnetic forces, the internal magnetic field configuration within the CME, and other expansion-driven effects. It is also possible that the presence of HSSS may partly contribute to the observed acceleration. Other factors may also play a role, including the possibility that SolO and Wind sampled slightly different parts of the CME due to their small longitudinal separation as discussed in \citeA{Regnault2024}. Overall, accurately modeling CME-solar wind interaction requires a more comprehensive analysis in which the drag parameter and the pre-conditioned background solar wind through which the CME propagates are reliably constrained.

Utilizing the estimated acceleration of each substructure and implementing the CAAP method to in situ observations (Equation~\ref{equ:speed}), the speed of each substructure at each instance on both spacecraft is estimated and is listed in normal font in the top panel of Table~\ref{tab:speed_tab}. The speeds listed in bold fonts are the in situ measured speed of the substructures at both spacecraft. By utilizing the propagation speeds of two substructures (from LE, center, and TE) at the same instance, the instantaneous expansion speed of the MC can be estimated at both spacecraft. We compute the instantaneous expansion speed at Wind on the arrival of the center ($t_2'$) and at SolO on the arrival of the TE ($t_3$). The bottom panel of the table lists the three expressions that can be used to estimate the instantaneous expansion speed of the MC at both spacecraft using the propagation speed of two substructures at the same instance.

From this panel, we note that the instantaneous expansion speed of the MC at both spacecraft, estimated from the expression containing the center and TE speed, matches well with the measured instantaneous expansion speed. However, the instantaneous expansion speed from the other two expressions is inconsistent with the measured instantaneous expansion speed; this could be because both expressions utilize the speed of the LE. Taking a constant positive acceleration of the MC LE for a long duration could result in overestimating its speed at time $t_2$ and $t_3$ ($t_2'$ and $t_3'$) for SolO (Wind), since the LE of the MC may experience strong drag force from the ambient medium in comparison to other CME substructures \cite{Vršnak2010,Mishra2013,Agarwal2024}. This shows that the CAAP method significantly relies on acceleration, and its accurate estimation would be required for the success of the method. In the current heliospheric era, where multiple spacecraft provide heliospheric imagers (HIs) and in situ measurements throughout the inner heliosphere, there would be a greater likelihood of estimating the acceleration of CME substructures for many more cases. With reliable acceleration estimates for CME substructures, the CAAP method can determine the instantaneous expansion speed even using the single-spacecraft in situ observations.

We also use the three expressions listed in the first column of the bottom panel in Table~\ref{tab:speed_tab} to estimate the CME time-independent expansion speed based on the conventional method. Therefore, the estimates of such time-independent expansion speed can be attributed to any time during the complete passage of the CME at the spacecraft. This method assumes that the propagation speed of each substructure remains constant during the complete passage of the whole CME over the in situ spacecraft. It would be interesting to compare the estimates of time-independent and time-dependent instantaneous expansion speeds from the conventional and non-conventional methods, respectively.

We note that among the three estimates of time-dependent expansion speeds from the CAAP method, the estimate from $V_C-V_T$ (second row in the table) is consistent with its measured value as described above. Therefore, we use this estimate for comparison with the time-independent expansion speed estimates. Using the same substructures, the time-independent expansion speeds at SolO and Wind are 43 and 33 $km~s^{-1}$, respectively, which are lower than the time-dependent expansion speed. However, this time-independent expansion speed accounts for the dynamics only in the second half of the CME radial width, while the expression $V_L-V_C$ accounts for the dynamics in the first half of the CME radial width. Therefore, we prefer to compare the time-independent expansion speed estimated using the expression $\frac{V_L - V_T}{2}$ (the third row expression), as this takes into account the dynamics of the CME during its complete (LE to TE) passage at the spacecraft. Earlier studies also utilized speed profiles from LE to TE to estimate the time-independent expansion speed \cite{Owens2005,Lugaz2020a,Zhuang2023}. The expansion speed accounting for the CME dynamics only in the first half or second half of the CME radial width may differ from one another, if there is a non-uniform speed profile before and after the CME center. From the fifth panel of Figure~\ref{fig:overplot}, we can notice the uniform speed profile for the SolO while a non-uniform speed profile for Wind, which is also reflected in the values in the table.

Based on our reliable estimates, the time-independent expansion speeds during the CME passage are 44 $km~s^{-1}$ and 50 $km~s^{-1}$ at SolO and Wind, respectively. While the value of time-dependent expansion speed is 67 $km~s^{-1}$ at the arrival of CME TE at SolO and its center at Wind. We note that the time-independent expansion speed is underestimated in comparison to time-dependent expansion speed estimates at both spacecraft. The underestimated value of time-independent expansion speed represents the lower estimates of CME's radial width and higher estimates of magnetic field strength. In Section~\ref{sec:flux_radius}, we discuss the influence of underestimated expansion speed on radial speed and magnetic field strength. We also present a qualitative analysis of MC's magnetic flux evolution from SolO to Wind. We further discuss the implications of underestimated expansion speed from the space weather perspective in Section~\ref{sec:resdis}. In the following section, we will examine the evolution of CME substructures from SolO to Wind.

\section{Evolution of CME Substructures: Shock, Sheath, and Magnetic Cloud} \label{sec:evocmesub}

In Section~\ref{sec:instexpa}, we found that the presence of HSSS at the rear portion of the MC at SolO is the possible reason for the acceleration of the MC substructures from SolO to Wind. Moreover, the presence of HSSS could have altered the evolution of CME substructures (shock, sheath, and MC) from their expected evolution. It will be interesting to examine the evolution of the selected CME substructures and find changes in them at SolO and Wind. In the following sections, we analyze the characteristics of the shock, sheath, and MC at both spacecraft. We also examine the evolution of the MC's axis from SolO to Wind by implementing the MVA technique on in situ observations of the MC.

\subsection{Shock Characteristics at Solar Orbiter and Wind} \label{sec:shock}

\begin{table}
\caption{The table lists the estimated shock parameters at SolO and Wind in the first and second row, respectively. The listed shock parameters are shock normal ($\hat{n}_{RTN}$) in the RTN coordinate system, shock speed ($V_{sh}$), ratio of downstream to upstream total magnetic field ($B_2/B_1$), velocity ($V_2/V_1$), density ($N_2/N_1$), and temperature ($T_2/T_1$). We also list the angle between the shock normal and the magnetic field ($\theta_{Bn1}$ for upstream and $\theta_{Bn2}$ for downstream), Alfvén Mach number ($M_{A1}$ for upstream and $M_{A2}$ for downstream), and fast magnetosonic Mach number ($M_{fms1}$ for upstream and $M_{fms2}$ for downstream).}

\begin{adjustbox}{width = 1.0\textwidth}
\centering
\begin{tabular}{ccccccccccccc}
\hline
\multicolumn{13}{c}{Shock Parameters} \\
\hline
{\multirow{2}{*}{Spacecraft}} & {Shock Normal} & {Shock Speed} & \multirow{2}{*}{$B_2/B_1$} & \multirow{2}{*}{$N_2/N_1$} & \multirow{2}{*}{$V_2/V_1$} & \multirow{2}{*}{$T_2/T_1$} & $\theta_{Bn1}$ & $\theta_{Bn2}$ & \multirow{2}{*}{$M_{A1}$} & \multirow{2}{*}{$M_{A2}$} & \multirow{2}{*}{$M_{fms1}$} & \multirow{2}{*}{$M_{fms2}$}\\ 
 & {($\hat{n}_{RTN}$)} & {{($V_{sh}$) [$km~s^{-1}$]}} &  &  &  &  & {$[^\circ]$} & {$[^\circ]$} &  &  &  & \\ 
\hline
{SolO} & {[0.60, -0.35, -0.72]} & {515} & {2.2} & {2.3} & {1.2} & {1.6} & {59} & {77} & {3.2} & {0.9} & {2.1} & {0.7} \\
{Wind} & {[0.83, 0.14, -0.55]} & {760} & {2.8} & {4.7} & {1.5} & {7.3} & {33} & {73} & {6.9} & {1.1} & {3.9} & {0.4} \\
\hline    
\end{tabular}
\label{tab:shock_tab}
\end{adjustbox}
\end{table}

We analyze the evolution of the shock from SolO to Wind by comparing shock parameters at both spacecraft. Due to a data gap in the SWE instrument at the time of the shock, we use 24-second resolution plasma measurements from the 3DP instrument to estimate shock parameters at Wind. The shock normal vector ($\hat{\mathbf{n}}$) at both spacecraft is determined in the RTN coordinate system using the Mixed Mode 3 (MX3) method \cite{Schwartz1998}, as described below:

\begin{equation*}
    \hat{\mathbf{n}}_{MX3} = \pm\frac{(\Delta \mathbf{B} \times \Delta \mathbf{V}) \times \Delta \mathbf{B}}{|(\Delta \mathbf{B} \times \Delta \mathbf{V}) \times \Delta \mathbf{B}|}
\end{equation*}

where $\pm$ sign of $\hat{\mathbf{n}}$ is arbitrary and can be chosen such that $\hat{\mathbf{n}}$ points in upstream of the shock, $\Delta \mathbf{B}$ is the difference between the magnetic field vector in downstream ($\mathbf{B_2}$) and upstream ($\mathbf{B_1}$) of the shock and, similarly, $\Delta \mathbf{V}$ is the difference between the velocity vector in downstream ($\mathbf{V_2}$) and upstream ($\mathbf{V_1}$) of the shock. The upstream and downstream states are represented by subscripts 1 and 2, respectively. The magnetic field and plasma parameters in the upstream and downstream of the shock are estimated by averaging the parameters over a fixed time interval. The averaging time window spans 10 minutes, excluding the 2-minute interval immediately before and after the shock. The estimated shock normal at both spacecraft is listed in the second column of Table~\ref{tab:shock_tab}. Further, the shock speed ($V_{sh}$) is estimated along the shock normal using the mass flux conservation equation as follows;

\begin{equation}\label{equ:shock_speed}
    V_{sh} = \left| \left( \frac{N_2\mathbf{V_2} - N_1\mathbf{V_1}}{N_2 - N_1} \right) \cdot \hat{\mathbf{n}} \right|
\end{equation}

where $N_2$ and $N_1$ represent the downstream and upstream density, respectively. Table~\ref{shock_tab} lists the estimated shock parameters at both spacecraft. The mathematical expressions used to estimate the other shock parameters are outlined in \citeA{Kilpua2015}. In addition to the jump in the magnetic field, density, temperature, and speed, we further checked whether the shock was only a sudden jump or a real fast-forward (FF) shock. For this purpose, we estimate the angle between the shock normal and the magnetic field vector as denoted by $\theta_{Bn1}$ for upstream and $\theta_{Bn2}$ for downstream of the shock. We also estimate the Alfvén Mach number ($M_A$) and fast magnetosonic Mach number ($M_{fms}$) in upstream and downstream of the shock. The evolutionary conditions for the FF shock are as follows: (i) $M_{fms1} > 1$, $M_{fms2} < 1$ (ii) $M_{A1} > 1$, $M_{A2} > 1$ (iii) $\theta_{Bn2} > \theta_{Bn1}$ \cite{Hsieh1986}. All estimated shock parameters are listed in Table~\ref{tab:shock_tab} at both spacecraft. We have considered an error of 15\% in these estimated shock parameters due to possible uncertainties in the plasma parameters and averaging/median over the chosen upstream and downstream region. The estimated value of $M_{A2}$ at SolO is 0.9 and does not strictly satisfy the criterion for an FF shock, but it falls within the corresponding error-bar range. Therefore, the observed discontinuities at both SolO and Wind can be confidently identified as an FF shock.

The shock speed at SolO is 515 $km~s^{-1}$, assuming it to be constant during its propagation from SolO to Wind, the expected arrival of the shock at Wind is after 10.3 hours. \citeA{Regnault2024}, the front speed of the sheath observed at SolO is used to infer that the shock would take roughly 8 hours to travel from SolO to Wind. However, the shock arrived in only half the time expected, as shown in the top panel of Table~\ref{tab:arr_rad_tab}. An increase in shock speed likely explains this, as the shock arrived at Wind with a speed of 760 $km~s^{-1}$. On comparing the compression ratios at both spacecraft, we note that the magnetic/density compression ratio at Wind (2.8/4.7) is more than its value at SolO (2.2/2.3). Moreover, the compression ratio of temperature at Wind is nearly 4.6 times its value at SolO. Further, the value of $M_{A1}$ and $M_{fms1}$ is higher at Wind, in comparison to their values at SolO, which is consistent with the increase in the shock compression ratio and shock speed at Wind. The estimates of the compression ratio and shock speed clearly suggest that the shock is stronger at Wind than at SolO. Such an acceleration and strengthening of the shock is not expected in general during the shock's outward journey. Contrary to our finding, the weakening of the shock is common and is understood due to its propagation in the expanding (lower-pressure) solar wind, energy dissipation, and a typical deceleration of the driving fast-speed CME. Earlier studies have noted deceleration of the shock during its journey farther out from the Sun \cite{Woo1985,Neugebauer2013}.

In our case study of the 3 November 2021 event, it is possible that some favorable solar wind conditions have caused the rarely observed increase in shock speed at a larger heliocentric distance. From Section~\ref{sec:instexpa}, we infer that the MC following the shock is undergoing acceleration, suggesting that the shock driver itself is accelerating. This acceleration could potentially explain the observed increase in shock speed at Wind. The increase in shock speed (Equation~\ref{equ:shock_speed}) at a larger distance from the Sun can also be understood by examining the relative reduction in mass flux ($(N_2\mathbf{V_2} - N_1\mathbf{V_1})\cdot \hat{\mathbf{n}}$) with density jump $(N_2 - N_1)$ during the propagation of the MC from SolO to Wind. We notice that both the mass flux and density jumps across the shock are reduced at Wind compared to SolO; however, the reduction in mass flux is less pronounced than that in density. The ratio of the jump in mass flux at Wind to that at SolO is approximately half, whereas the corresponding ratio for the jump in density is about one-third. This could clearly result in a favorable condition to increase the shock speed at Wind. Further, the inherent or evolutionary changes in the shock geometry locally or geometrical effects due to the spacecraft's location relative to the shock front could be possible reasons for the difference in the shock speed at SolO and Wind.

The strengthening of the shock is also possible if the driver propagates into a background medium of reduced Alfvén speed. We observe a decrease in Alfvén speed from SolO to Wind, primarily due to a larger reduction in the upstream magnetic field than in the square root of the upstream density. The values of $M_{A1}$ and $M_{fms1}$ at Wind are higher than those at SolO, as there is an increase in the solar wind velocity in the shock frame by a factor of $\sim$ 1.5 and a decrease in the Alfvén speed at Wind by a factor of $\sim$ 1.4. Further, the higher value of $M_{fms1}$ at Wind than that at SolO is due to a decrease in the value of $\theta_{Bn1}$ as the shock propagated from SolO to Wind. The value of $\theta_{Bn}$ in the downstream of the shock is higher than its value in the upstream of the shock at both spacecraft. Notably, the downstream-to-upstream difference of $\theta_{Bn}$ is significantly higher at Wind than at SolO. This indicates a stronger compression in the downstream of the shock at Wind. Our analysis of shock and CME characteristics shows that the shock is strengthened during its propagation from SolO to farther out up to Wind. Our case study shows the possibility of a shock amplification with increasing distance from the Sun.

\subsection{Sheath Characteristics at Solar Orbiter and Wind} \label{sec:sheath}

In Figure~\ref{fig:insitu_solo_wind}, the transparent red region depicts the CME sheath, with durations of 17.1 and 16.5 hours at SolO and Wind, respectively. We estimate the sheath radial size at SolO and Wind by utilizing Equation~\ref{equ:radialsize} to examine the ongoing variations in the sheath between both locations. The estimated radial size of the sheath at SolO and Wind is 61.9 $R_\odot$ and 61.3 $R_\odot$, respectively. These values are listed in the bottom panel of Table~\ref{tab:arr_rad_tab}. We find that the sheath duration and radial size at the two spacecraft are nearly identical, which is unexpected given the CME’s propagation from SolO to Wind over a radial distance of 0.13 AU. One would anticipate a larger sheath at Wind due to the continued pile-up of upstream solar wind plasma between SolO and Wind. Additionally, an increase in shock speed from SolO to Wind could lead to enhanced compression, while the acceleration of the MC over this interval could further contribute to sheath size.

Shock-MC interactions with upstream solar wind are expected to imprint on the sheath properties at both spacecraft. The sheath's average speed and temperature are slightly higher at Wind than at SolO, consistent with compression by an accelerating shock and MC. A combination of factors, related to shock, MC, and ambient medium, can lead to constancy of sheath size with increasing distance from the Sun. The sheath size is found to have no clear correlation with speed and Mach number of the MCs \cite{Salman2020a,Temmer2022}. The study of \citeA{Mishra2021a} also shows the average sheath size to be constant for CMEs in solar cycles 23 and 24, despite CMEs having an average speed higher in cycle 23. The size of the accumulated sheath seems to be a time-integrated effect reaching its saturation within a certain distance, and future study is required to confirm this.

\subsection{Magnetic Cloud Characteristics at Solar Orbiter and Wind} \label{sec:mc}

In Figure~\ref{fig:insitu_solo_wind}, the transparent yellow region represents the MC, with durations of 12.5 and 16.1 hours at SolO and Wind, respectively. Unlike the nearly similar duration of the sheath at both spacecraft, the duration of the MC at Wind is 3.6 hours longer than that at SolO. The estimated radial size of the MC at the arrival of the MC LE at SolO and Wind is 39.3 $R_\odot$ and 53.5 $R_\odot$, respectively. From this, the MC size is noticeably larger at Wind in comparison to its size at SolO, while the radial size of the sheath at both spacecraft is nearly the same. This implies that the MC expands from SolO to Wind. The presence of HSSS at the rear portion of the MC at SolO (as shown in the fifth panel of the left side panel of Figure~\ref{fig:insitu_solo_wind}) could have compressed and accelerated the MC. The effect of MC's compression, confirmed at SolO, and possibly continued for some time, could be a reason for its increased speed and temperature at Wind.

The top panel of Table~\ref{tab:arr_rad_tab} presents the arrival times of MC substructures (LE, center, and TE) at SolO and Wind, along with the differences between them. \citeA{Regnault2024} assumed that if the LE had propagated between SolO to Wind at the constant speed that was measured at SolO (a distance of 0.13 AU), it would have arrived at Wind approximately 8 hours after its passage at SolO. If we apply the same constant‐speed assumption to the center and TE, the expected arrival times of center and TE at Wind would have been approximately 8.6 and 9.3 hours, respectively, after their passage at SolO. However, as indicated in the fourth column of the top panel of Table~\ref{tab:arr_rad_tab}, the actual arrival times of LE, center, and TE at Wind are significantly earlier than estimates from the constant-speed assumption. This discrepancy suggests that the MC substructures underwent acceleration during their propagation from SolO to Wind, as found from propagation speed measurements at both spacecraft and discussed in Section~\ref{sec:instexpa}. Our analysis extends the constant-speed assumption approach to other substructures of center and TE, and results are consistent with earlier studies \cite{Regnault2024,Rossi2025}.

The orientation of the MC's axis is an important characteristic of the flux rope from the space weather perspective \cite{Mulligan1998}. We visually determine the orientation of the selected MC by analyzing the rotation of the latitude and longitude of the total magnetic field vector within the MC boundary (as shown in Figure~\ref{fig:insitu_solo_wind}), which is East-North-West (ENW) at both spacecraft \cite{Bothmer1998}. This implies that the selected MC is observed as a highly inclined flux rope \cite{Mulligan1998,Nieves-Chinchilla2019} at both spacecraft. We estimate the inclination ($\theta$)/azimuthal angle ($\phi$) of the MC axis from/within the ecliptic plane by utilizing the MVA technique \cite{Sonnerup1967,Bothmer1998,Agarwal2025}.

The estimated parameters from the MVA technique at SolO and Wind are listed in Table~\ref{tab:tab_2}. The first row of the table lists the maximum ($\lambda_1$), intermediate ($\lambda_2$), and minimum ($\lambda_3$) eigenvalues of the magnetic variance matrix (detailed description of the magnetic variance matrix is explained in \citeA{Agarwal2025}). The measured magnetic field vector projected into the direction of the three eigenvectors ($e_1$, $e_2$, and $e_3$ corresponding to their eigenvalues) provides the magnetic field components $B_x^*$, $B_y^*$, $B_z^*$ in the direction of maximum, intermediate, and minimum variances, respectively. These estimated directions of variances are well determined if $\lambda_2/\lambda_3\ge2$ \cite{Siscoe1972,Lepping1980}. For the selected MC, the values of $\lambda_2/\lambda_3$ at SolO and Wind are 3.3 and 2.1, respectively, as listed in the second row of the table. This ensures that the variances are well determined for the selected 3 November 2021 MC, and therefore, the direction of the MC axis derived from MVA is reliable at both SolO and Wind spacecraft to understand the MC evolution.

\begin{table}
\caption{The table lists the estimated parameters from the MVA technique at SolO and Wind in the second and third columns, respectively.}
\centering
\begin{tabular}{ccc}
\hline
\multicolumn{3}{c}{MVA Results} \\
\hline
{Parameters} & {SolO} & {Wind} \\
\hline
    {Eigenvalues ($\lambda_1$, $\lambda_2$, $\lambda_3$)} & {(25.9, 3.0, 0.9)} & {(21.9, 2.3, 1.1)}\\
    {$\lambda_2/\lambda_3$} & {3.3} & {2.1} \\ 
    {Intermediate Eigenvector ($e_2$)} & {(0.99, -0.04, 0.12)} & {(0.9, 0.12, 0.41)} \\
    {Minimum Eigenvector ($e_3$)} & {(0.12, 0.03, -0.99)} & {(0.37, 0.28, -0.89)} \\{Maximum Eigenvector ($e_1$)} & {(0.04, 0.99, 0.03)} & {(-0.22, 0.95, 0.21)} \\
    {Orientation ($\theta$, $\phi$) of the MC Axis} & {(7$^\circ$, 357.7$^\circ$)} & {(24.3$^\circ$, 7.3$^\circ$)}\\
    {Orientation ($\theta$, $\phi$) of the $e_3$} & {(-82.7$^\circ$, 13.8$^\circ$)} & {(-62.4$^\circ$, 37.7$^\circ$)}\\
    {Orientation ($\theta$, $\phi$) of the $e_1$} & {(2$^\circ$, 87.9$^\circ$)} & {(12.4$^\circ$, 103$^\circ$)}\\
\hline
\end{tabular}
\label{tab:tab_2}
\end{table}

\begin{figure*}
    \centering
    \includegraphics[scale = 0.7,trim={0cm 0cm 0cm 0cm},clip]{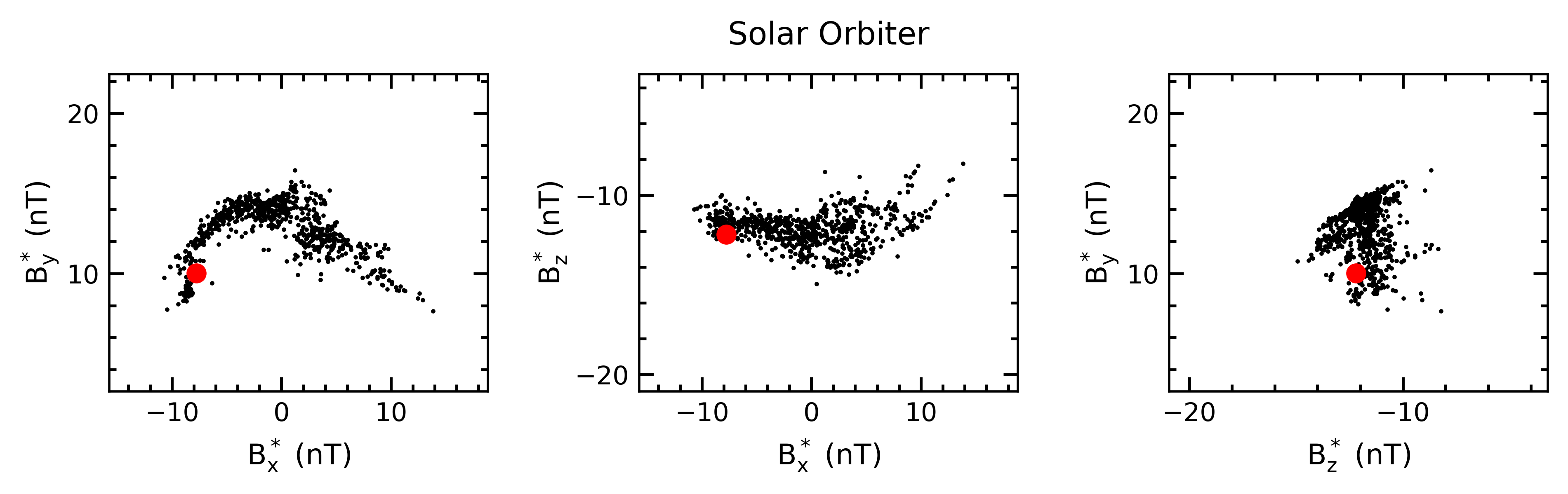}
    \includegraphics[scale = 0.7,trim={0cm 0cm 0cm 0cm},clip]{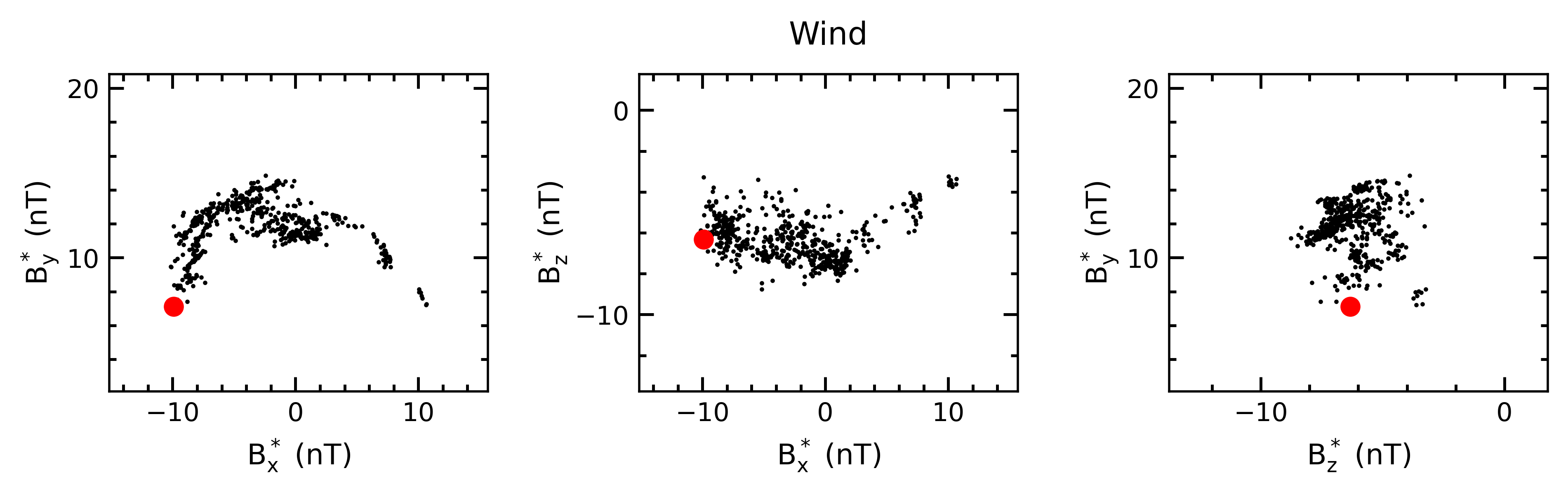}
    \caption{Hodogram representation between $B_x^*$, $B_y^*$, and $B_z^*$ at SolO and Wind in the top and bottom panels, respectively. The measured magnetic field vectors in the direction of maximum, intermediate, and minimum variances are $B_x^*$, $B_y^*$, and $B_z^*$. The red dot represents the starting point of the MC.}
    \label{fig:hodo}
\end{figure*}

Moreover, the measured magnetic field vectors projected into the direction of maximum, intermediate, and minimum variances become $B_x^*$, $B_y^*$, and $B_z^*$ components. Figure~\ref{fig:hodo} depicts the hodogram representation, after corrected for ambiguity, between any two magnetic field components, $B_i^*$ (where $i = (x, y, z)$), for SolO in the top panel and for Wind in the bottom panel. The ambiguity in MVA arises because the eigenvalues ($\lambda_1$, $\lambda_2$, and $\lambda_3$) of the magnetic variance matrix are the actual variances in $B_x^*$, $B_y^*$, and $B_z^*$ and are always positive. Therefore, they can correspond to both positive and negative ($e_i$ and $-e_i$) eigenvectors as valid directions of variances. Initially, the hodogram representation of $B_x^*$ with $B_y^*$ indicated for a WNE-type MC at both spacecraft; however, the in situ observations suggest for ENW. Therefore, to ensure the consistency of the rotation of $B_x^*$ with $B_y^*$, we changed the sign of both eigenvectors $e_1$ and $e_3$ to make the right-hand coordinate system. The red dot represents the starting point of the MC. The consistent eigenvectors at both spacecraft from MVA with in situ measurements are listed in Table~\ref{tab:tab_2}.

Figure~\ref{fig:hodo} illustrates the hodogram representation consistent with in situ measurements at both spacecraft. The hodogram between $B_y^*$ and $B_x^*$ shows a clear rotation in $B_x^*$ from negative to positive as the value of $B_y^*$ changes while maintaining its positive sign at both spacecraft. The second hodogram between $B_z^*$ and $B_x^*$ shows the change in the value of $B_x^*$ from negative to positive with a slight variation in the value of $B_z^*$ at both spacecraft; the variation in $B_z^*$ is more pronounced at Wind. The third hodogram between $B_y^*$ and $B_z^*$ shows the change in the value of $B_y^*$ while maintaining its positive sign, with the nearly constant value of $B_z^*$ at both spacecraft. Further, we have estimated the direction/orientation of the maximum, intermediate, and minimum eigenvectors using the expressions described in \citeA{Agarwal2025}.

The estimated orientations ($\theta$, $\phi$) of the intermediate, minimum, and maximum eigenvectors at both spacecraft are listed in the last three rows of Table~\ref{tab:tab_2}. According to \citeA{Bothmer1998}, the directions of the intermediate and minimum eigenvectors correspond to the MC axis and the MC propagation direction, respectively. The orientation of the MC axis ($\theta$, $\phi$) at SolO and Wind is found to be (7$^\circ$, 357.7$^\circ$) and (24.3$^\circ$, 7.3$^\circ$), respectively. These values indicate that the MC axis is generally aligned with the Sun-Earth line at both spacecraft. While the axis lies nearly within the ecliptic plane at SolO, it is slightly inclined out of the ecliptic at Wind. Overall, this suggests that the MC axis is not highly inclined at either location.

Although MVA-derived results suggest a low inclination of the flux rope at both spacecraft, the visual inspection of in situ observed magnetic field vectors indicates an ENW-type flux rope at both locations, which is an attribute of high inclination. This apparent discrepancy between the MVA-derived flux rope orientation and the visually inspected flux rope type could arise if the MC propagation direction is significantly out of the in situ spacecraft plane. This will ensure that the spacecraft intersects the MC along an oblique path, and away from its central axis. In such cases where in situ measurements are away from the central axis of the flux rope, the intermediate component in MVA cannot be reliably used to estimate the orientation of the MC axis. Furthermore, the MVA-derived propagation directions ($\theta$, $\phi$) are (-82.7$^\circ$, 13.8$^\circ$) at SolO and (-62.4$^\circ$, 37.7$^\circ$) at Wind, suggesting an unrealistically out-of-ecliptic propagation, which is again inconsistent with the well-observed MC signatures at both spacecraft. In this context, \citeA{Li2022} reported the propagation direction of the selected CME as N20E09, which is estimated by utilizing the Graduated Cylindrical Shell (GCS) model near the Sun. This highlights a substantial difference between the MVA-derived propagation direction and the GCS-based result. Since MVA is known to be sensitive to the selected MC boundaries, we tested this effect on our selected MC by shifting both the LE and TE boundaries by $\pm$10 minutes and found only a change within $\pm$10$^\circ$ for the derived axis orientation and propagation direction. This shows that the MVA results are not highly sensitive to boundary selection for the selected CME. The source of these discrepancies, such as between MVA-derived axis orientation and visually identified flux-rope type, MVA-derived propagation direction, and GCS-based results, remains unclear; further investigation using additional modelling is required in future studies.

\subsubsection{Influence of Expansion Speed on Radial Size and Magnetic Flux of Magnetic Cloud}\label{sec:flux_radius}

The expansion of MC during its outward journey from the Sun causes its radial size ($r$) to increase with heliocentric distance ($h$). The evolving radial size typically follows a power-law relation $r\propto h^{n_r}$, where $n_r$ is the power-law index. Earlier studies have estimated this index by estimating the MC's size using the propagation speed of the MC across its duration at spacecraft located at different heliocentric distances, which lies in the range $0.45<n_r<1.14$. \cite{Bothmer1994,Bothmer1998,Gulisano2010,Leitner2007,Dumbovic2018}.

$$n_r = \frac{\log_e(r_2/r_1)}{\log_e(h_2/h_1)}$$

where $r_2$ represents the radius of the MC at the heliocentric distance of the MC's LE, denoted as $h_2$; $r_1$ corresponds to the radius of the MC at an earlier heliocentric distance $h_1$ \cite{Dumbovic2018,Lugaz2020a,Zhuang2023}. One can also investigate the power law index by examining the evolution of radial size with heliocentric distance using the expansion speed of the MC at different distances, through the relation $r_2 = r_1 + \int_{t_1}^{t_2} V_{exp}(t)dt$; where $V_{exp} (t)$ represents the instantaneous expansion speed. In this expression, one can assume that size is precisely measured at initial height $h_1$. In Section~\ref{sec:instexpa}, we show that the estimated expansion speed from the conventional method is time-independent (non-instantaneous) during the CME passage at the in situ spacecraft and is underestimated for the selected case compared to the time-dependent (instantaneous) expansion speed. This suggests that using the time-independent expansion speed from the conventional approach at $h_1$ to infer the MC radial size at a larger heliocentric distance $h_2$ would underestimate the MC size at that location. Consequently, the power-law index ($n_r$) describing the radial-size evolution with distance will also be underestimated.

Such underestimation will be larger for fast-accelerating CMEs compared to slowly-accelerating CMEs, assuming the actual radial sizes were the same in both cases. Likewise, for decelerating CMEs, the expansion speed from the conventional method will be overestimated, and if this estimate at a particular distance is assumed constant to find the radial size at later distances, the power law governing the radial size will also be overestimated. This discussion emphasizes the need to use the instantaneous expansion speed at different times during the passage of the MC at the spacecraft. In contrast, if the propagation speed of the MC center and the expansion speed remain constant (i.e., no acceleration or deceleration), then the expansion speed from the conventional method is consistent with the instantaneous expansion speed.

Further, the expansion governs the magnetic field strength dilution inside the MC, considering the conservation of the magnetic flux ($\phi \propto B_c r^2$), where $B_c$ is the magnetic field strength of the MC axis/center and $r$ is the radius of the MC \cite{Dumbovic2018}. The underestimated time-independent expansion speed, if extrapolated forward for higher distances, will lead to an underestimated radial size, which in turn results in an overestimation of $B_c$. It is suggested that there is a power-law relation between the magnetic field of the MC and its distance from the Sun, $B\propto h^{-n_B}$, where $n_B$ is the power-law index \cite{Dumbovic2018}. The power-law indices for radius and central magnetic field strength are as follows:

$$n_B = \frac{-\log_e(B_2/B_1)}{\log_e(h_2/h_1)}$$

where $B_2$ represents the central magnetic field strength of the MC at the heliocentric distance of the MC's LE, denoted as $h_2$; while $B_1$ corresponds to the central magnetic field strength of the MC at an earlier heliocentric distance $h_1$ \cite{Dumbovic2018,Lugaz2020a,Zhuang2023}. \citeA{Dumbovic2018} highlighted that the magnetic flux of an MC is conserved if $n_B = 2n_r$, while an increase or decrease in magnetic flux with heliocentric distance corresponds to $n_B < 2n_r$ or $n_B > 2n_r$, respectively.

The selected MC evolved from the SolO at $h_1 = 0.85$ AU to Wind at $h_2 = 0.98$ AU, and it would be interesting to examine the magnetic flux ($\phi$) evolution between both spacecraft. We can express $\phi_2/\phi_1 = (h_2/h_1)^{-(n_B - 2n_r)}$, where $\phi_1$ and $\phi_2$ are the magnetic flux at SolO and  Wind, respectively. We note that the measured values of the central magnetic field/radius of the MC are 18.9 nT/19.7 $R_\odot$ and 15 nT/26.8 $R_\odot$ at SolO and Wind, respectively. These estimates provide the power-law indexes as $n_B = 1.6$ and $n_r = 2.2$. This shows that the rate of increase in radial size is greater than the rate of decrease in magnetic field strength.

Moreover, under the force-free flux rope assumption, the poloidal magnetic flux of a flux rope can be estimated using the central magnetic field strength ($B_c$), the flux‐rope radius ($r$), and the length of the MC axis ($l$), where the axis length scales with heliocentric distance ($h$) \cite{Wang2015}. Using these relations, we find that the poloidal magnetic flux of the MC increases by about 24\% during its propagation from SolO to Wind. Additionally, for the selected MC, \citeA{Zhang2025} found a 13\% increase in the axial magnetic flux. Together, these results suggest that neither of the magnetic flux components is conserved between the two locations, suggesting additional flux injection from SolO to Wind. Previous studies have suggested that magnetic reconnection at the TE of the MC can lead to such additional flux injection \cite{Manchester2014,Dumbovic2018}. Possible signatures of magnetic reconnection are discussed in detail in Section~\ref{sec:resdis}. Throughout this analysis, we have assumed that the axial magnetic field corresponds to the estimates at the size center, which is valid only if MC’s axial center coincides with the size center \cite{Agarwal2025}. This assumption cannot be strictly valid for the selected MC, as it exhibits signatures of distortion at the rear edge. However, it does not significantly affect our analysis of poloidal flux, as the magnetic field strength remains nearly uniform throughout the MC passage at both spacecraft.

\section{Summary and Discussion}\label{sec:resdis}

The present study focuses on the validation of our proposed non-conventional method, or Constant Acceleration Accounted Perspective (CAAP), for the single-point in situ observations for estimating the time-dependent instantaneous expansion speed of the MC. The validation is done on a selected CME that was observed by radially-aligned SolO and Wind spacecraft--during 3-4 November 2021 at SolO and at Wind during 3-5 November 2021. This MC is quite unique and rare as its center is observed at Wind, and at the same time, its TE is observed at SolO. The simultaneous propagation speed measurements of the MC center at Wind and the TE at SolO provide the measured instantaneous expansion speed of the MC as 68 $km~s^{-1}$. The measured instantaneous expansion speed is found to be in good agreement with estimated values from our CAAP (non-conventional) method applied independently to single-point SolO and Wind observations. Our study demonstrates the usefulness and advantages of the CAAP method over the conventional method for estimating expansion speeds, while underscoring that its accuracy depends on reliable input values of constant acceleration of CME substructures.

In a recent study by \citeA{Regnault2024}, the instantaneous expansion speed of the selected MC was measured to be 73 $km~s^{-1}$. The slight discrepancy between their estimate and ours arises from the differing definitions of the CME center. While \citeA{Regnault2024} used the time center--that is, the midpoint of the CME’s temporal profile--as the reference point for calculating expansion, our study adopts the size center, which represents the midpoint in spatial extent between the leading and trailing edges. The instantaneous expansion speeds estimated using the CAAP method may vary slightly depending on the assumed acceleration; the conventional method significantly underestimates the expansion speed and, as a result, the radial size of the CME. Such underestimations can significantly affect the accuracy of space weather forecasting, as several key parameters--such as the arrival time of the CME, the duration of geomagnetic disturbances, and the geoeffectiveness--are influenced by the CME's expansion.

If the estimated instantaneous expansion speed (67 $km~s^{-1}$) is assumed to be constant, the radial size of the MC should have increased by only about $2~R_\odot$ during its LE arrival from SolO to Wind. However, the actual radial size of the MC from SolO to Wind is found to increase by $\sim 7~R_\odot$ at the arrival of the LE. This implies that the non-constant instantaneous expansion speed exists between SolO and Wind. In this context, our CAAP method with reliable estimation of the CME substructure's acceleration can offer a promising alternative for deriving the time-dependent instantaneous expansion speed of CMEs.

The study also focuses on analyzing the properties of a selected CME's substructures (shock, sheath, and MC) at two radially separated locations. The radial size of the MC at Wind is nearly 1.4 times its value at SolO, potentially due to its continued expansion between SolO and Wind. The expansion at SolO might have been prevented to some extent due to the observed presence of HSSS and its compressing effect at the rear side of the MC. The arrival time of MC substructures (LE, center, and TE) at Wind from SolO is earlier than their predicted arrival time, especially for the LE and center. This can be the result of the observed acceleration of MC substructures due to the momentum transfer from HSSS to CME \cite{Gopalswamy2009a}. Given the small angular separation between SolO and Wind for the selected event, the observed differences in the MC properties at the two spacecraft are more likely attributable to the temporal evolution of the MC and its propagation through a non-linear, anisotropic ambient medium, and less likely due to sampling different regions of the structure.

We also note a slight difference in the MC axis orientation (higher at Wind) from the MVA technique at both spacecraft, which is consistent with previous studies showing the possibility of variations in MC axis orientation even for small angular or radial spacecraft separations \cite{Liu2008,EmmaDavies2020,Agarwal2025}. The finding of higher tilt at later heights at Wind than at SolO contrasts with the findings of \citeA{Zhang2025}, which report a decrease in MC axis tilt with increasing heliocentric distance. Additionally, an ENW-type flux rope is inferred from the visual inspection of in situ measurements, which attributes a highly inclined structure inconsistent with the MVA-derived axis orientations. This discrepancy could arise if the MC had propagated out of the spacecraft plane and/or if the spacecraft sampled the structure away from its center along an oblique trajectory. In this case, the intermediate eigenvector from MVA may not reliably represent the MC axis. For this event, the MVA-derived propagation directions at both spacecraft lie significantly out of the ecliptic plane, in strong disagreement with the GCS-based direction reported by \citeA{Li2022}. Furthermore, the ratio of intermediate to the minimum eigenvalue is only slightly above the threshold criteria for MVA reliability; the possibility of large uncertainties in the MVA results cannot be ignored. A more comprehensive investigation, incorporating additional remote-sensing observations and modelling, is required to better interpret the evolution of the CME axis and propagation directions in the IP medium.

In our study, the observed strengthening of the shock at Wind is interesting, as in general, the shock is expected to weaken during its journey in the IP medium \cite{Woo1985,Neugebauer2013}. The strengthening of the shock is possible due to the observed acceleration of the MC as the driver of the shock. The studies utilizing observations from multi-spacecraft, with longitudinal and radial separation even as small as 0.03 AU and 4$^\circ$, respectively, have suggested that the shape of the shock could be non-spherical and shock parameters could be different along the shock front under the influence of non-isotropic medium or due to asymmetry in the shock driver's properties \cite{Mostl2012,Kilpua2021,Mishra2021,Lugaz2022,Palmerio2024}. Therefore, the role of the shock driver and the ambient medium seems crucial for evolving shock properties in our case, and further, the role of a smaller longitudinal separation between the spacecraft responsible for sampling the differing local structure at the shock front cannot be fully ignored.

In our study, the estimated shock parameters at both spacecraft are nearly equal to their values cataloged in the Heliospheric Shock Database (\url{https://ipshocks.helsinki.fi/database}). However, the estimated shock speed, magnetic field compression ratio, density compression ratio, $\theta_{Bn1}$, upstream Alfvén Mach number, and upstream fast magnetosonic Mach number at SolO are significantly different from the values reported in \citeA{Trotta2023}. This discrepancy could be possible due to the use of different averaging windows (ranging from 30 seconds to 5 minutes) in their study to estimate the upstream/downstream parameters. This suggests that the estimated shock parameters are highly sensitive to the averaging window used for upstream/downstream parameters. The shock examined in our study appears to be relatively strong, as its estimated parameters (shock speed, upstream Alfvén Mach number, magnetic field, density, and temperature compression ratio) exceed the median values reported for a large sample of shocks in \citeA{PerezAlanis2023}.

We note that CME's sheath size and its duration are nearly identical at both spacecraft. The sheath properties are known to be influenced by a combination of factors related to the shock, the driver of the shock, and the ambient medium \cite{Salman2020a,Kilpua2021,Temmer2022}. Given that the factors influencing sheath size differ between SolO and Wind, their combined effects may cancel out, suggesting that the observed sheath size was already established before the event reached SolO. The sheath size and duration of the selected event are about 2-3 times larger than the median value of sheath size and duration over a large sample of CMEs at 1 AU \cite{Winslow2015,Janvier2019,Salman2020,Temmer2022}. The higher duration/size of the sheath, compared to its average value in the literature, could be possible if the spacecraft has cut the selected CME/MC away from its center, which agrees with the results of the MVA technique at both spacecraft. It is evident that understanding the formation of the sheath is not straightforward, as its size cannot be directly attributed to the properties of the shock drivers or the ambient solar wind.

For the selected MC, we find that magnetic flux is injected during its propagation from SolO to Wind, using the estimated power-law indexes $n_B$ and $n_r$. We note that the estimates of the $n_B$ take the central magnetic field strength at a later time than the time at which the radius of the MC is used. The measured magnetic field of the MC's center at the same time as the MC LE arrival would have been larger. This suggests that indeed there is a flux injection from SolO to Wind for the selected MC in our study, which is in contrast to some of the earlier studies \cite{Owens2006,Dasso2006,Ruffenach2012}. The possibility of flux injection due to magnetic reconnection at the TE of the MC is also reported earlier \cite{Manchester2014,Dumbovic2018}. For the selected MC, at both spacecraft, we note the plausible signatures of magnetic reconnection at the TE of the MC, a dip in the magnetic field along with an increase in proton velocity, density, temperature, and plasma beta \cite{Gosling2005,Remeshan2026}. A more definitive confirmation would require examining electron pitch-angle distributions (PADs); however, for this event, the reconnection signatures are not sufficiently clear to draw a conclusive inference. However, it remains to be confirmed whether the HSSS at the rear of the MC, and the associated compression and heating observed at SolO, enable the MC to utilize thermal pressure more effectively than magnetic pressure for its expansion. This could potentially account for the observed increase in size accompanied by a relatively smaller decrease in magnetic field strength at Wind.

In the study of \citeA{Zhuang2023}, they suggest for having excluded the effect of expansion speed from the propagation speeds to estimate the MC's radial size by using the expression $(V_L - V_{exp})\Delta t$; where $V_L$, $V_{exp},~\mathrm{and} ~\Delta t$ are the LE speed, expansion speed from the conventional method, and total duration of the MC passage at the spacecraft, respectively. The expression $V_L - V_{exp}$ is exactly the propagation speed of the MC's time center (equally divides the MC's total duration into two parts), and their expression can estimate the MC’s size at the arrival of the LE, provided the MC exhibits a linearly decreasing speed-time profile throughout its passage at the spacecraft. We note that the method of \citeA{Zhuang2023} ignores the effect of the propagation speed of all the individual substructures within the LE to TE; the estimated size of the MC can be an overestimate or an underestimate of that from the trapezoidal rule (Equation~\ref{equ:radialsize}) adopted in our study. Assuming a case of strictly expanding MC, i.e., observed LE speed is greater than the TE speed, the method of \citeA{Zhuang2023} will overestimate the size if the MC expands more in the initial segment than its the later segment, while it will underestimate if the MC expands less (relatively compressed) in the initial segment than in the later segment. These effects could also arise due to the aging effects, such as acceleration/deceleration, deflection, and distortion, acting on the CME substructures following the LE over the in situ spacecraft. Moreover, if the measured propagation speed at the size center is considered over the entire MC duration with a linearly decreasing speed-time profile, it will provide an overestimated radial size than that from its actual value (using the trapezoidal rule) and that from using the method of \citeA{Zhuang2023}. This occurs because the spacecraft takes unequal amounts of time to pass through the first and second halves of the MC's radial size.

It is obvious that measured in situ propagation speeds involve the contribution of expansion speeds, even if there is no change in the propagation speed of the MC center during its complete passage over the spacecraft. The approach of expansion correction by \citeA{Zhuang2023} is valid if there is no bulk acceleration/deceleration in the MC. Therefore, the method of deriving expansion-corrected propagation speeds requires generalization applicable to any speed-time profile and at any instant during the MC passage over the spacecraft. The general expression for expansion corrected speed of any substructures (at any time) within an expanding MC would be $V_p' (t) = V_p (t) \mp \frac{r(t)}{R(t)}V_{exp}(t)$, where $V_p(t)$ is the measured in situ propagation speed, and $V_p'(t)$ is the expansion corrected propagation speed at time $t$. The $R(t)$ and $V_{exp}(t)$ are the instantaneous MC's radius and expansion speed at any time $t$. The negative sign in the equation is for substructures before the size center, while the positive sign is for substructures after the size center. For such generalization, the instantaneous expansion speed can be inferred using remote observations, MHD modeling, and/or simultaneous measurements of CME substructures from radially aligned in situ spacecraft, which does not necessarily have to be derived using the CAAP method. This expansion corrected propagation speed excludes the contribution of expansion speed and can thus be used to infer the true acceleration or deceleration of the MC center or bulk acceleration of the MC.

The CAAP method depends on the assumption of a constant acceleration of CME substructures during their passage over an in situ spacecraft. The acceleration estimates derived from the propagation speeds of CME substructures--either using radially aligned in situ spacecraft or remote-sensing observations--can be subject to significant uncertainties if the CME undergoes rotation or deflection during its propagation. In such cases, radially aligned spacecraft may sample different CME substructures, leading to errors in the inferred acceleration. Similarly, acceleration estimates from remote observations may not correspond to the same substructures later sampled by in situ spacecraft, thereby introducing additional uncertainty in the derived instantaneous expansion speed from the CAAP method. Moreover, if CME distortion or interaction with large-scale solar wind structures occurs beyond the heliocentric distance at which the acceleration has been estimated--whether from in situ or remote-sensing observations--then incorporating such acceleration into the CAAP method may introduce additional uncertainties in the inferred instantaneous expansion speed. The presence of complex in situ speed profiles also suggests that expansion may be non-uniform along the spacecraft trajectory, which challenges the assumption of constant acceleration of CME substructures in the CAAP method. Despite these limitations, the CAAP method (based on its concept) remains promising in the modern heliospheric era, where multiple spacecraft equipped with HIs and in situ instruments can provide enhanced observational coverage of CMEs in the IP medium. Our study also underscores the need and importance of placing in situ monitors at various sub-L1 distances. These observations can provide improved constraints on the acceleration of CME substructures in the inner heliosphere, enabling a more reliable estimation of instantaneous expansion speeds.

Several MHD models use the CME’s 3D kinematics near 0.1 AU as an initial boundary condition for the IP evolution of CMEs \cite{Odstrcil2003,Pomoell2018,Barnard2022,Mayank2024,Owens2025}. These models represent the CME cross-section using various geometries--such as circular, elliptical, spheromak, or flux-rope configurations--these geometries, undergoing evolution in background solar wind conditions, can inherently constrain the CME’s instantaneous expansion speed. There are limited attempts to validate the model-derived instantaneous expansion speed with the in situ observed instantaneous expansion speed. If there exists a mismatch between model and observed expansion speeds, it may be possible that the predicted CME LE arrival time is reasonably accurate, while the modeled radial size, duration, magnetic flux content, and the arrival times or speeds of individual substructures can still deviate significantly from observations. Therefore, it is important that future studies make an attempt to compare the instantaneous expansion speed from models and observations, for which the selected CME offers a valuable test case.

Our study finds that the selected MC exhibits unexpected evolution from SolO to Wind, including accelerated CME substructures, an increase in shock speed, a nearly constant sheath size, enhanced magnetic flux, and significant discrepancies between MVA-derived orientations and visually identified flux‐rope characteristics. These results highlight the complex nature of MC evolution in the IP medium, suggesting that key physical processes governing such evolution warrant further in-depth studies. The particular event also demonstrates that the CME behavior at L1 can remain difficult to reliably predict, even when sampled in advance by radially aligned in situ spacecraft. To address these challenges, future studies should account for the temporal and spatial variability of the non-isotropic ambient solar wind and internal evolution of magnetized CMEs as factors to govern the evolution of CMEs. Such external and internal factors can possibly explain the unusual acceleration/deceleration of CMEs as discussed in \citeA{Rossi2025}. Therefore, detailed analysis of similar events is essential for improving current modeling capabilities and enhancing predictions of CME substructure arrival times, radial sizes, and geoeffectiveness.

The present study validates our CAAP method, in contrast to the conventional method, for estimating the time-dependent instantaneous expansion speed of CMEs using single-point in situ observations. The method is validated on a single case study, using in situ observations independently of SolO and Wind. Since there exist statistical studies using the conventional method, it is required to use the CAAP method over several CMEs to further confirm its accuracy and possible use in space weather forecasting techniques using future sub-L1 observations. The differences in the properties of MC at SolO and Wind are understood as the combined effect of the time-evolution of the substructures, non-isotropic ambient medium, and the spacecraft relative separation from the CME center and its propagation direction. Future studies on the statistics of such CMEs using more than two multipoint spacecraft can shed light on the processes causing changes in CME properties at multiple locations in the heliosphere.

\section*{Conflict of Interest}

The authors declare that there are no conflicts of interest relevant to this study.

\section*{Open Research}

The Solar Orbiter and Wind spacecraft data used in this study are publicly available from the NASA Coordinated Data Analysis Web (CDAWeb) at \url{https://cdaweb.gsfc.nasa.gov/}. The data sources and their documentation for Solar Orbiter are provided by ESA at \url{https://www.cosmos.esa.int/web/soar/instrument-documentation}, and those for Wind are provided by NASA at \url{https://wind.nasa.gov/data.php}. All plots in this study are produced using open-source Python 3.9.18 packages, including NumPy, Matplotlib, cdflib, and datetime.

\acknowledgments
This research was funded in whole or in part by the Austrian Science Fund (FWF) [10.55776/P36093]. For open access purposes, the author has applied a CC BY public copyright license to any author-accepted manuscript version arising from this submission.

\end{document}